\documentclass[a4paper,fleqn,usenatbib,useAMS]{mnras}

\usepackage{newtxtext,newtxmath}

\usepackage[T1]{fontenc}

\DeclareRobustCommand{\VAN}[3]{#2}
\let\VANthebibliography\thebibliography
\def\thebibliography{\DeclareRobustCommand{\VAN}[3]{##3}\VANthebibliography}

\usepackage{graphicx}	
\usepackage{subfigure}
\usepackage{natbib}
\usepackage{color}
\usepackage{tabularx}
\usepackage{longtable}
\usepackage{bm}
\usepackage{threeparttable}

\usepackage[normalem]{ulem}

\newcommand{\tess}{{\it TESS}}

\newcommand{\tar}{HD 21749}
\newcommand{\can}{HD 21749c}
\newcommand{\gaia}{{\it Gaia}}
\newcommand{\jwst}{{\it JWST}}
\newcommand{\elt}{{\it ELT}}

\title[\tar b, c]{Revisiting the HD 21749 Planetary System with Stellar Activity Modeling\thanks{This paper includes data gathered with the 6.5 meter Magellan Telescopes located at Las Campanas Observatory, Chile.}}

\author[T. Gan et al.]{Tianjun Gan,$^{1}$\thanks{Corresponding authors: gtj18@mails.tsinghua.edu.cn}
Sharon~Xuesong~Wang,$^{1,2}$\thanks{Corresponding authors: sharonw@mail.tsinghua.edu.cn}
Johanna~K.~Teske,$^{2,\ast}$
Shude~Mao,$^{1,3}$
Ward S. Howard,$^{4}$
\newauthor
Nicholas M. Law,$^{4}$
Natasha~E. Batalha,$^{5}$
Andrew~Vanderburg,$^{6,\S}$
Diana~Dragomir,$^{7}$
Chelsea~ X.~Huang,$^{8,\star}$
\newauthor
Fabo Feng,$^{9}$
R.~Paul~Butler,$^{9}$
Jeffrey~D.~Crane,$^{2}$
Stephen~A.~Shectman,$^{2}$
Yuri Beletsky,$^{2}$
\newauthor
Avi~Shporer,$^{8}$
Benjamin~T.~Montet,$^{10}$
Jennifer A. Burt,$^{11}$
Adina~D.~Feinstein,$^{12,\P}$
Erin Flowers,$^{13,\P}$
\newauthor
Sangeetha Nandakumar,$^{14}$
Mauro Barbieri,$^{14}$
Hank Corbett,$^{4}$
Jeffrey K. Ratzloff,$^{4}$
Nathan Galliher,$^{4}$
\newauthor
Ramses Gonzalez Chavez,$^{4}$
Alan Vasquez,$^{4}$
Amy Glazier,$^{4}$
Joshua Haislip$^{4}$
\\
$^{1}$Department of Astronomy and Tsinghua Centre for Astrophysics, Tsinghua University, Beijing 100084, People's Republic of China\\
$^{2}$Observatories of the Carnegie Institution for Science, 813 Santa Barbara Street, Pasadena, CA 91101, USA\\
$^{3}$National Astronomical Observatories, Chinese Academy of Sciences, 20A Datun Road, Chaoyang District, Beijing 100012, People's Republic of China\\
$^{4}$Department of Physics and Astronomy, University of North Carolina at Chapel Hill, Chapel Hill, NC 27599-3255, USA\\
$^{5}$NASA Ames Research Center, Moffett Field, CA 94035, USA\\
$^{6}$Department of Astronomy, University of Wisconsin-Madison, Madison, WI 53706, USA\\
$^{7}$Department of Physics and Astronomy, University of New Mexico, 1919 Lomas Blvd NE, Albuquerque, NM 87131, USA\\
$^{8}$Department of Physics and Kavli Institute for Astrophysics and Space Research, Massachusetts Institute of Technology, Cambridge, MA 02139, USA\\
$^{9}$Earth \& Planets Laboratory, Carnegie Institution of Washington 5241 Broad Branch Road, N.W., Washington, DC 20015, USA\\
$^{10}$School of Physics, University of New South Wales, Sydney, NSW 2052, Australia\\
$^{11}$Jet Propulsion Laboratory, California Institute of Technology, 4800 Oak Grove Drive, Pasadena, CA 91109, USA\\
$^{12}$Department of Astronomy and Astrophysics, University of Chicago, 5640 S. Ellis Ave, Chicago, IL 60637, USA\\
$^{13}$Department of Astrophysical Sciences, Princeton University, Princeton, NJ 08544, USA\\
$^{14}$INCT, Universidad De Atacama, calle Copayapu 485, Copiap\'{o}, Atacama, Chile\\
$^{\ast}$NASA Hubble Fellow\\
$^{\S}$NASA Sagan Fellow\\
$^{\star}$Juan Carlos Torres Fellow\\
$^{\P}$NSF Graduate Research Fellow
}

\date{Accepted XXX. Received YYY; in original form ZZZ}

\pubyear{2020}

\begin{document}
\label{firstpage}
\pagerange{\pageref{firstpage}--\pageref{lastpage}}
\maketitle

\begin{abstract}
\tar\ is a bright ($V=8.1$ mag) K dwarf at 16~pc known to host an inner terrestrial planet \can\ as well as an outer sub-Neptune \tar b, both delivered by TESS. Follow-up spectroscopic observations measured the mass of \tar b to be $22.7\pm2.2\ M_{\oplus}$ with a density of $7.0^{+1.6}_{-1.3}$ g~cm$^{-3}$, making it one of the densest sub-Neptunes. However, the mass measurement was suspected to be influenced by stellar rotation. Here we present new high-cadence PFS RV data to disentangle the stellar activity signal from the planetary signal. We find that \tar\ has a similar rotational timescale as the planet's orbital period, and the amplitude of the planetary orbital RV signal is estimated to be similar to that of the stellar activity signal. We perform Gaussian Process (GP) regression on the photometry and RVs from HARPS and PFS to model the stellar activity signal. Our new models reveal that \tar b has a radius of $2.86\pm0.20\ R_{\oplus}$, an orbital period of $35.6133\pm0.0005$ d with a mass of $M_{b}=20.0\pm2.7\ M_{\oplus}$ and a density of $4.8^{+2.0}_{-1.4}$ g~cm$^{-3}$ on an eccentric orbit with $e=0.16\pm0.06$, which is consistent with the most recent values published for this system. \can\ has an orbital period of $7.7902\pm0.0006$ d, a radius of $1.13\pm0.10\ R_{\oplus}$, and a 3$\sigma$ mass upper limit of $3.5\ M_{\oplus}$. Our Monte Carlo simulations confirm that without properly taking stellar activity signals into account, the mass measurement of \tar b is likely to arrive at a significantly underestimated error bar. 
\end{abstract}




\section{Introduction}
\par After the first discovery of a transiting planet HD 209458b \citep{Henry2000,Charbonneau2000}, the field of exoplanet research entered a new chapter. The success of ground based transit surveys such as HATNet \citep{Bakos2004}, SuperWASP \citep{Pollacco2006} and KELT \citep{Pepper2007,Pepper2012} led to the detections of many giant planets \citep{Collier2007,Penev2013}. Later space missions including {\it CoRoT} \citep{Baglin2006}, {\it Kepler} \citep{Borucki2010} and {\it K2} \citep{Howell2014} produced an increasing number of terrestrial planets with well measured radii which in turn enabled the detection of a bimodal distribution of planet radii occurrence rate \citep{Fulton2017}. Using measurements from the radial velocity method that constrains planet masses, planet densities and compositions can be further studied (e.g., \citealt{Rogers2015}). 

Even though more than 4000 exoplanets\footnote{ \url{https://exoplanetarchive.ipac.caltech.edu/}} have been detected, most of them are too faint for RV follow up observations to determine their masses. The recently launched Transiting Exoplanet Survey Satellite (\tess, \citealt{Ricker2014,Ricker2015}) could increase the population of exoplanets with well measured radius and mass significantly, as \tess\ focuses on nearby bright stars\ ($V<13\ {\rm mag}$) and performs an all-sky survey. During the primary mission, \tess\ is expected to discover $\sim 10^4$ planets \citep{Sullivan2015,Huang2018}, and more than 2000 candidates have been identified, which is already providing the community with a large sample of ideal targets for further RV mass measurement and atmospheric characterization with the upcoming {\it James Webb Space Telescope} (\jwst, \citealt{Gardner2006}) and {\it Extremely Large Telescope} (\elt, \citealt{Gilmozzi2007,Zeeuw2014}).

Accurate and precise mass measurement is critical for analyzing the atmospheric profiles and elemental abundances via transmission or emission spectroscopy \citep[e.g.,][]{Batalha2019} as well as for modeling planetary interior structures in some cases \citep[e.g.,][]{Otegi2020}. For example, the equation below shows the degeneracy between the planet mass $M_{p}$ and the mean molecular weight $\rm \mu$ \citep{Seager2000,Seager2009}:
\begin{equation}
    H = \frac{kT}{\mu \times {\rm g_{p}}} \propto \frac{1}{\mu M_{p}},
\label{eqn:atmosphere}
\end{equation}
where $\rm g_{p}$ represents the surface gravity, and $H$ and $T$ are the scale height and temperature of the atmosphere, respectively. Indeed, the work by \cite{Batalha2019} found that, for warm Neptunes and Earth-sized planets, the quality of the measured planetary masses heavily dictates the accuracy and precision of the retrieved atmospheric parameters such as temperature and scale height.

However, stellar activity, often manifesting as stellar rotation signals, can cause large RV variations that pose challenges to accurate mass determination \citep{Queloz2009,Howard2013,Pepe2013}. Stellar activity may thus make it particularly challenging to characterize small planets ($R < 4R_{\oplus}$), whose RV signals are often on par with or even smaller than the signals induced by stellar activity. In recent years, Gaussian process (GP) regression has been widely used to account for the stellar activity signals in RVs, and GP regression has demonstrated success in teasing out the signals of small planets with RV amplitudes down to an order of magnitude smaller than the stellar activity \citep[e.g.,][]{Grunblatt2015,Rajpaul2015,Morales2016}.

The case of HD 21749 represents a somewhat extreme entanglement between stellar activity and planetary signals. This planetary system was first reported by \cite{Trifonov2019} and \cite{Dragomir2019} (here after D19). The rotation period of the star around 30--40~d is very close to the planet's orbital period near 35~d. Previous successful cases of disentanglement using GP very often dealt with systems where the two signals operate on very different timescales, typically with $P_{\rm planet}$ significantly smaller than $P_{\rm rot}$ \citep[e.g.,][]{Haywood2014,Grunblatt2015,Morales2016,Dai2017}. While D19 does not expect the RV signal of \tar b to be strongly affected by stellar variability, our new analyses in this work show that, very inconveniently, the RV amplitude of the stellar activity signal is probably similar to that of the planet. Here we present an updated planet mass measurement of \tar b where we disentangle the planetary component from the stellar activity signal in the RVs. 

\par This paper is organized as follows: In Section \ref{obs}, we describe the key observational data on \tar\ used in this work, including \tess\ photometry as well as HARPS and PFS RVs. We analyze all data at hand in Section \ref{analysis} including a study of the stellar rotation and our modeling of the stellar activity and planetary signals. In Section \ref{simulation}, we use our Monte Carlo simulations to evaluate the robustness of our model. Section \ref{discussion} presents our discussions about the prospects of atmospheric characterization of \tar b, including a search for additional planets. We conclude our findings in Section \ref{conclusions}.

\section{Observations}\label{obs}
The \tess\ photometry and PFS archival RVs in this work are the same as used in D19, except that we choose to adopt the HARPS data from the RVBANK \citep{Trifonov2020}. We direct the readers to \cite{Trifonov2019} and D19 for a description of the HARPS observations. Here we briefly describe the \tess\ photometry in Section \ref{tessphoto} and the PFS observations in Section \ref{pfs}. We have collected 147 more RV data points with PFS since D19, and so our final RV data set consists of 283 RV points.


\subsection{TESS Photometry}\label{tessphoto}


\par The original two-minute images of \tar\ were reduced by using the Science Processing Operations Center (SPOC) pipeline \citep{Jenkins2016} developed at NASA Ames Research Center. Based on the results from Transit Planet Search \citep[{TPS;}][]{Jenkins2002,Jenkins2017}, \tar\ was alerted on the MIT TESS Alerts portal\footnote{\url{https://tess.mit.edu/alerts/}} as a single-transit planet candidate (TOI 186.01). After recovering a second partial transit of \tar b near a momentum dump in the Sector 3 data and detecting a third full transit in Sector 4, D19 successfully confirmed the planetary nature of \tar b when combining \tess\ data with the observations from the High Accuracy Radial velocity Planet Searcher (HARPS, \citealt{Mayor2003}) and the Planet Finder Spectrograph (PFS, \citealt{Crane2006,Crane2008,Crane2010}). Furthermore, D19 also validated the inner Earth-sized planet \can\ (TOI 186.02) based on the data from previous three \tess\ Sectors [1,2,3]. The \tess\ light curve used in our analysis was extracted from the publicly available target pixel stamps using a simple photometric aperture with a customized pipeline (the same as the one in D19), which has not been corrected for the dilution or any instrumental systematic signals (see the third paragraph of section 2.1 in D19) 



\par We use the Box Least Square algorithm (BLS, \citealt{Kovacs2002}) to search for additional potential transiting planets after smoothing the light curve with a median filter. We confirm the 35.6 day and 7.8 day signals reported in D19. After masking out all these in-transit data, we do not find any other significant transit signals that exist in the current \tess\ data. 

\subsection{Evryscope-South}
Evryscope-South, which is located at Cerro Tololo Inter-American Observatory in Chile \citep{Law2015}, also observed 29059 epochs of HD 21749 over the course of 2.46 years. The two Evryscope facilities (North and South) are comprised of arrays of small telescopes that continuously image the entire accessible sky each night down to an airmass of two. Each facility has an instantaneous footprint of 8150 square degrees and a total coverage of 18,400 square degrees. Evryscope-South employs a ``ratchet" strategy that tracks the sky for 2 hours before ratcheting back into the initial position and continuing observations. The system observes at 2 min cadence in \textit{g}\textsuperscript{$\prime$}~\citep{Evryscope2015} at a pixel resolution of 13\arcsec pixel$^{-1}$ for $\sim$6 hr each night. The typical dark-sky limiting magnitude is \textit{g}\textsuperscript{$\prime$}=16. Astrometry and background subtraction are performed for each 28.8 MPix image. Raw photometry is extracted with forced apertures at target coordinates, then differential photometry is applied as described in \cite{Ratzloff2019}.

\subsection{PFS High Resolution Spectroscopy}\label{pfs}

We observed \tar\ with PFS on the 6.5~m Magellan II Clay telescope at Las Campanas Observatory in Chile starting in January 2010, about three months after PFS's first light, as part of the PFS long-term survey program to search for planets around nearby stars (e.g., \citealt{Teske2016,Feng2019}). The work in D19 includes RVs from PFS taken until December 14, 2018 (BJD 2458476.6).

PFS is an environmentally-controlled high-resolution spectrograph that uses an iodine cell for calibration to produce high precision RVs. The spectral reduction and RV extraction for PFS data are performed with a custom IDL pipeline that is capable of delivering RVs with $<$1~m/s precision \citep{Butler1996,Butler2019}. PFS was upgraded in January 2018 with a large format CCD, and the default RV observations were switched to a mode with a narrower slit (0.3\arcsec; pre-upgrade mode was with a 0.5\arcsec slit). This increased the spectral resolution from $R \sim 80,000$ to $R \sim 130,000$ and led to an increase in RV precision. We denote the RV observations taken with the pre-upgrade PFS with ``PFSF" (before BJD 2458000) and the post-upgrade ones with ``PFSS".

Starting October 2018, \tar\ was observed as one of the targets in the Magellan \tess\ Survey (MTS), which will characterize $\sim$30 \tess\ discovered super-Earths and sub-Neptunes to assemble a statistically robust sample of small planets (Teske et al. in prep.). Since D19, we have obtained 147 additional RV observations (53 epochs) as part of MTS. 

The 42 PFSF RV observations were taken with a typical exposure time of 300 seconds, mostly one exposure per epoch (41 epochs), and a set of template observations (3$\times$550-second exposures) were taken through the 0.3\arcsec slit with the pre-upgrade PFS without using the iodine cell. The 181 PFSS observations (62 epochs) consist of mostly pairs of two $\sim$300-second exposures, sometimes followed by another pair of observations a few hours later within the same night. This intra-night high cadence was designed to average out stellar jitter \citep{Dumusque2011} and potential instrumental noise. In addition, the MTS observations on \tar\ were executed with high cadence ($\sim$nightly visit when possible) in order to achieve good sampling on both the timescale of the stellar rotation and that of the planetary orbit. 

We also derive two spectroscopic activity indicators from the PFS data: the emission flux in the Ca \textsc{II} H \& K lines ($S_{\rm HK}$; \citealt{Duncan1991}) and the H$\alpha$ line ($S_{\rm H\alpha}$; \citealt{Gomesdasilva2011}), both residing in spectral regions free of iodine absorption lines. We perform the barycentric correction using Precise EXOplanetology (PEXO, \citealt{pexo}).  The PFS data used in this work are publicly available on ExoFOP-TESS.\footnote{\url{https://exofop.ipac.caltech.edu/tess/target.php?id=279741379}} 


\section{Analyses and Results}\label{analysis}
In this work, we adopt the stellar radius $R_{\star}=0.695\pm0.030$ $R_{\odot}$, mass M$_{\star}=0.73\pm0.07$ M$_{\odot}$, and effective temperature $T_{\rm eff}=4640\pm100$ K from D19. We encourage readers to refer to D19 for more details.

\subsection{Stellar Rotation}\label{rotation}

\begin{figure}
\centering
\includegraphics[width=0.5\textwidth]{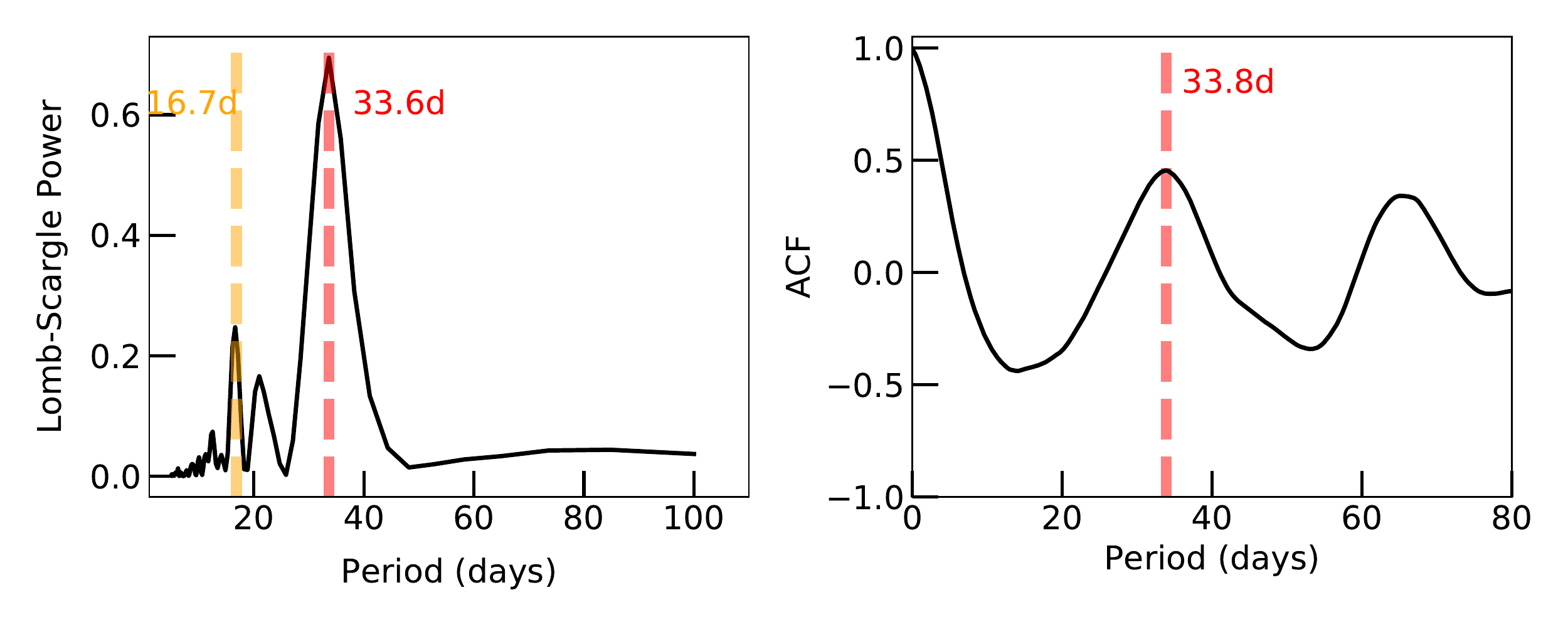}
\caption{{\it Left panel:} Lomb-Scargle (LS) power spectrum of the systematics-corrected \tess\ light curve with the peaks marked by vertical dashed lines. {\it Right panel:} Autocorrelation function\ (ACF) of the same dataset. Both of them show a consistent rotation period $\sim$ 33\ d (see Section \ref{rotation}).} 
\label{LA_rotation}
\end{figure}

We first evaluate the evidence for the stellar rotation signal and determine the rotation period of \tar. We employ the Lomb-Scargle\ (LS) periodogram to perform the frequency analysis \citep{Lomb1976,Scargle1982} and search for rotational spot modulation in the \tess\ photometry (Sector 1-Sector 4; baseline $\sim 110$~d). For this analysis, we exclude the second segment of data in Sector 4 where it exhibits a sharp downturn between BJD = 2458420 and BJD = 2458424 (see the blue highlighted part in Figure \ref{transit+GP}), possibly due to a series of events on board \tess\ including a camera shut-off and a heater turn-on, causing the PSF to change. We explore the period space between 5\ d to 100\ d after masking out all transits. We find a prominent peak at $\sim$ 33.6~d (red dashed line in Figure \ref{LA_rotation}) which is close to the values reported in D19. A less significant peak located at $\sim$16.7\ d, nearly half of the rotation period, is likely the second harmonic of the rotation signal (orange dashed line in Figure \ref{LA_rotation}). Following \cite{Gan2020}, we compute the auto-correlation function\ (ACF) of the \tess\ light curve using Numpy.correlate \citep{McQuillan2013}. The ACF exhibits a broad peak at $\sim$33.8\ d, which again suggests the $\sim$ 33\ d photometric signal could be the real rotation period of \tar\ (see Figure \ref{LA_rotation}). We note that although there are systematic signals in the extracted \tess\ light curve (e.g., due to reaction wheel momentum dump events; \citealt{Vanderburg2019}) and an elevated noise in Sector 4, it is unlikely to have a relatively smooth systematic signal with such a long period ($\sim$33~d) and large amplitude (200--400~ppm). Therefore, we conclude that the $\sim$33~d signal is due to the stellar rotation.

A recent study from \cite{Martins2020} reported a 66.8 d  rotation period using Section 1--3 of the \tess\ photometry spanning 88~d. This periodicity of 66.8~d is about twice the period of our finding, and we think it is a multiple of the true rotation period near 33~d. First, with an effective temperature of $T_{\rm eff}=4640\pm100$~K and a stellar age of $3.8\pm3.7$ yrs (D19), the mid-K dwarf \tar\ is highly unlikely to have a rotation period  $P_{\rm rot}>60$~d according to gyrochronology (e.g., see Figure 5 in \citealt{McQuillan2014}\footnote{\cite{McQuillan2014} used data from Kepler, where periods longer than  $\sim 30$ days were suppressed by PDC.}; \citealt{Barnes2017,Mamajek2008,Curtis2019}). Moreover, we detect a strong periodic signal around 31~d in the PFS magnetic activity indicators $S_{\rm HK}$ and $S_{\rm H\alpha}$ (see below), which suggests that \tar\ is probably not a very old and slowly rotating star. Thus we conclude that this 66.8 d period is the 2$P_{\rm rot}$ signal.

\begin{figure}
	\centering
	{
		\includegraphics[width=3.4in]{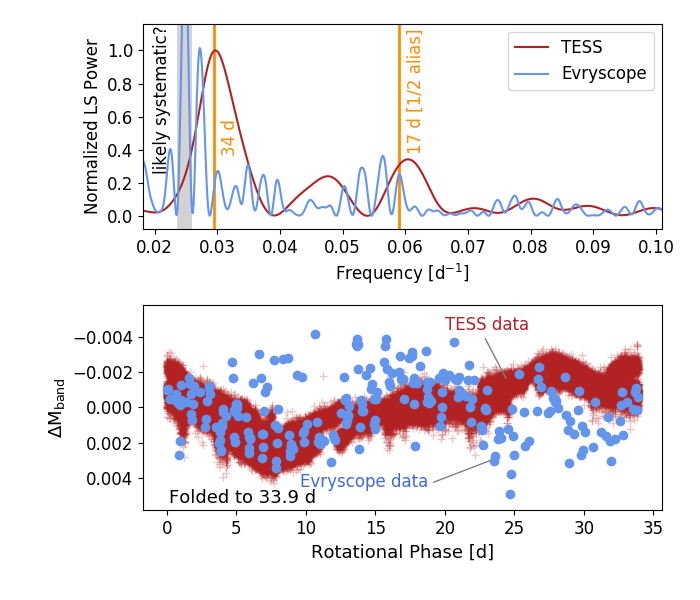}
	}
	\caption{\textit{Top panel:} Evryscope and \tess\ LS periodograms show power near 17 d. \tess\ also shows power at twice this period, $\sim$34 d. A likely systematic period at 40.3 d is highlighted in grey. \textit{Bottom panel:} The Evryscope and \tess\ light curves are phase-folded at a period of 33.9 d. The \tess\ amplitude is $\sim$2 mmag, close to the detection limit of phase-folded Evryscope observations. Evryscope observations are binned in phase to increase the detectability. Variability of the same phase and amplitude as in \tess\ is observed at low signal to noise.}
	\label{evryscope_figure}
\end{figure}

Additional ground based photometry also supports this finding. We observe periodic variability at $\sim$33.9~d (blue points in Figure~\ref{evryscope_figure}) with observations by Evryscope-South. We pre-whiten the light curve, iteratively removing likely systematics-affected periods including periods at 1 d, 51 d, 90 d, and 365 d. We then phase-fold the light curve to 33.9 d. Due to the low $\sim$2 mmag amplitude, we bin the light curve in phase to increase the signal to noise. The Evryscope periodogram displays a series of peaks near the 1/2 alias of the $\sim$33 d \tess\ signal but does not detect the $\sim$33 d signal directly. Multiplying the clearest Evryscope signal at 16.97 d by two, we phase-fold the light curve at 33.9 d and observe a signal of similar amplitude and phase as in the \tess\ light curve as shown in Figure \ref{evryscope_figure}. Other nearby periods do not match the \tess\ rotational phase as closely as the 16.97 d signal. We note that HD 21749 has a $g^{\prime}$ magnitude close to the non-linear regime of Evryscope. Furthermore, strong daily and yearly cycles are also present in the light curve prior to pre-whitening. These factors make our detection tentative but part of a cumulative case for a $\sim$33-34 d period. The 40.3 d signal is likely an Evryscope systematic as it is similar to spurious periods seen in other light curves and does not correctly phase up the \tess\ light curve. More detail on the Evryscope light curve of HD 21749 is given in Howard et al. 2020, in preparation.

Using photometric data from KELT \citep{Pepper2007,Pepper2012}, \cite{Oelkers2018} derived a rotation period of 47.6~d for \tar, while D19 reported the tallest peak in the Lomb-Scargle periodogram is at 38.95~d with the second tallest peak near 33~d.  The relatively low precision and systematic noise, which commonly associated with ground-based photometry, prevent us from detecting the rotation period of \tar\ in a statistically significant way. We prefer the $P_{\rm rot}$ value derived from the space photometry as \tess\ data has much higher precision. 

\par
\begin{figure}
\centering
\includegraphics[width=0.5\textwidth]{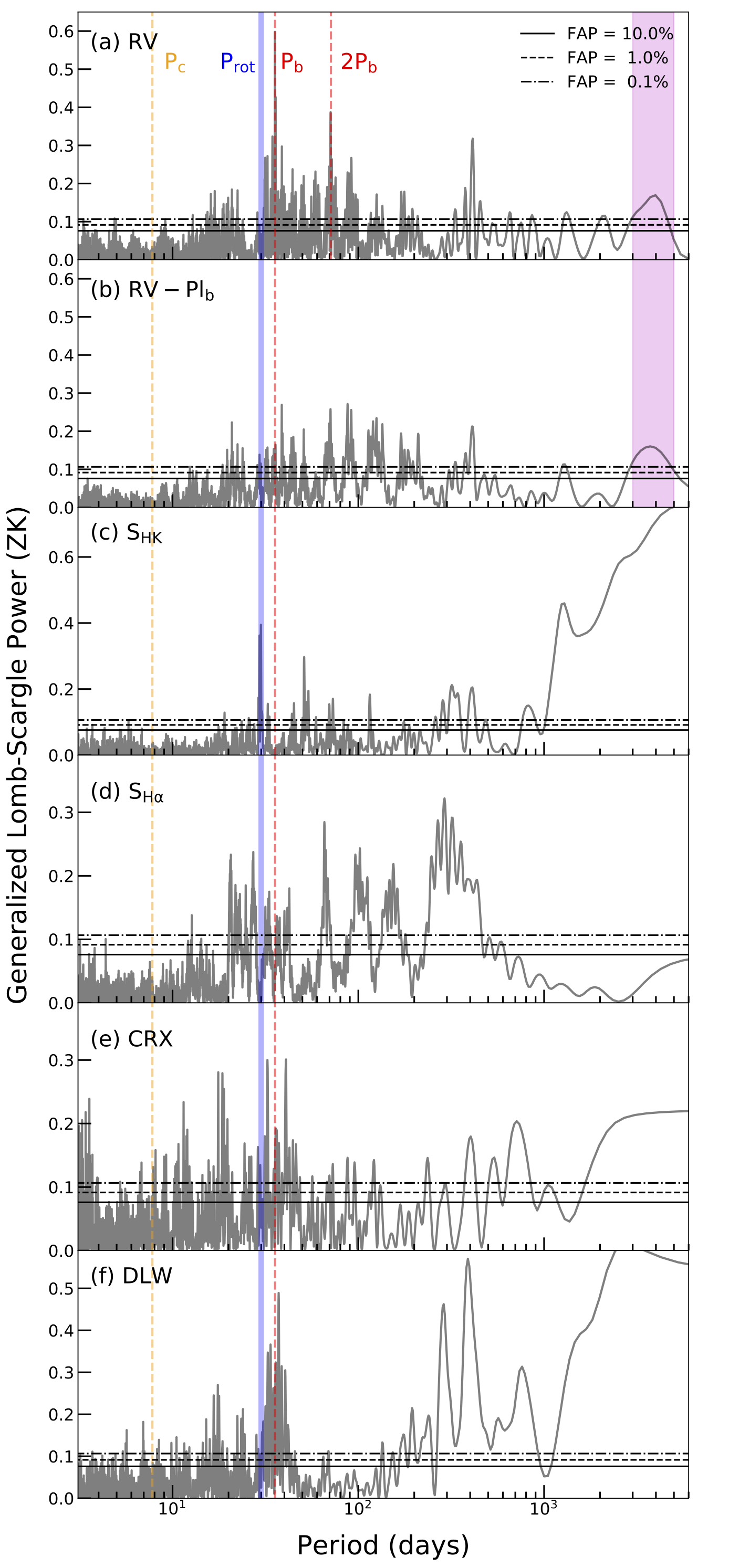}
\caption{GLS periodograms of: (a). Combined HARPS and PFS radial velocity data (after adjusting for the RV offsets between different instruments using the best-fit values in our GP+1pl model presented in Section~\ref{rv-only}). (b). Combined HARPS and PFS radial velocity data after subtracting the best-fit Keplerian model of \tar b. (c-d). PFS stellar activity indicators $S_{\rm HK}$ and $S_{\rm H\alpha}$. (e-f). HARPS stellar activity indicators CRX and DLW. The orbital periods of \tar b and the planet candidate \can\ are shown as red and orange vertical dashed lines, respectively. The shaded vertical region represent the signal of a potential outer planet companion (see Section \ref{add_pl}). The stellar rotation period 31\ d derived from the spectroscopy data is shown as blue vertical lines. The theoretical FAP levels of 10\%, 1\% and 0.1\% are marked as horizontal solid, dashed and dot-dashed lines.}
\label{gls_spec}
\end{figure}

 We then derive the rotation period from the spectroscopic data. D19 constrained the rotation period to be $34.5\pm7$~d based on the log($R_{\rm HK}$) index \citep{Mamajek2008}, and $\sim37$~d based on the LS periodograms of HARPS activity indicators ($S_{\rm HK}$ and $S_{\rm H\alpha}$). Additionally, the rotation speed ($v\sin i\sim$ 1.04 km/s) obtained from the HARPS spectra places a constraint on $P_{\rm rot}/$ $\sin i$ around 33.8\ d (D19). We search for periodic signals in the combined HARPS and PFS radial velocity data by utilizing the generalized Lomb-Scargle\ (GLS) periodogram \citep{Zechmeister2009}. A highest peak around 35.6\ d is detected at the orbital period of the transiting planet \tar b. After subtracting the optimized Keplerian solution (described below), we do not find any other significant peaks. We further investigate the GLS periodogram of the stellar activity indicators: $S_{\rm HK}$ and $S_{\rm H\alpha}$ indices from PFS, as well as the chromatic RV index\ (CRX) and the differential line width\ (DLW) from HARPS. The activity indicators from the HARPS spectra were extracted by using the publicly available SpEctrum Radial Velocity AnaLyser pipeline\ (SERVAL, \citealt{Zechmeister2018,Trifonov2020}). \cite{Trifonov2019} mentioned that the strongest power in DLW is at lower frequencies, which is also true for $S_{\rm H\alpha}$ and $S_{\rm HK}$. After excluding an outlier ($S_{\rm H\alpha}=-1.0$), we find an obvious peak at around 31\ d in the $S_{\rm HK}$ periodogram and a forest of peaks around 30\ d in the $S_{\rm H\alpha}$ periodogram , which adds credence to the $P_{\rm rot}$ value inferred in the \tess\ light curve analysis above. However, we do not find a significant peak in HARPS RV data that is consistent with the $\sim$ 31\ d rotation signal seen in $S_{\rm HK}$. This may be due to the relatively poor sampling on the timescale of the stellar rotation. We show all our frequency analyses in Figure \ref{gls_spec}. Both photometric and spectroscopic evidences show that $P_{\rm rot}$ is between 31 d and 37 d. Thus we choose to place a relatively wide uniform prior ([30, 40]) on $P_{\rm rot}$ in the following GP+transit modeling (see Section \ref{photometry-only}).

\begin{figure}
\centering
\includegraphics[width=0.5\textwidth]{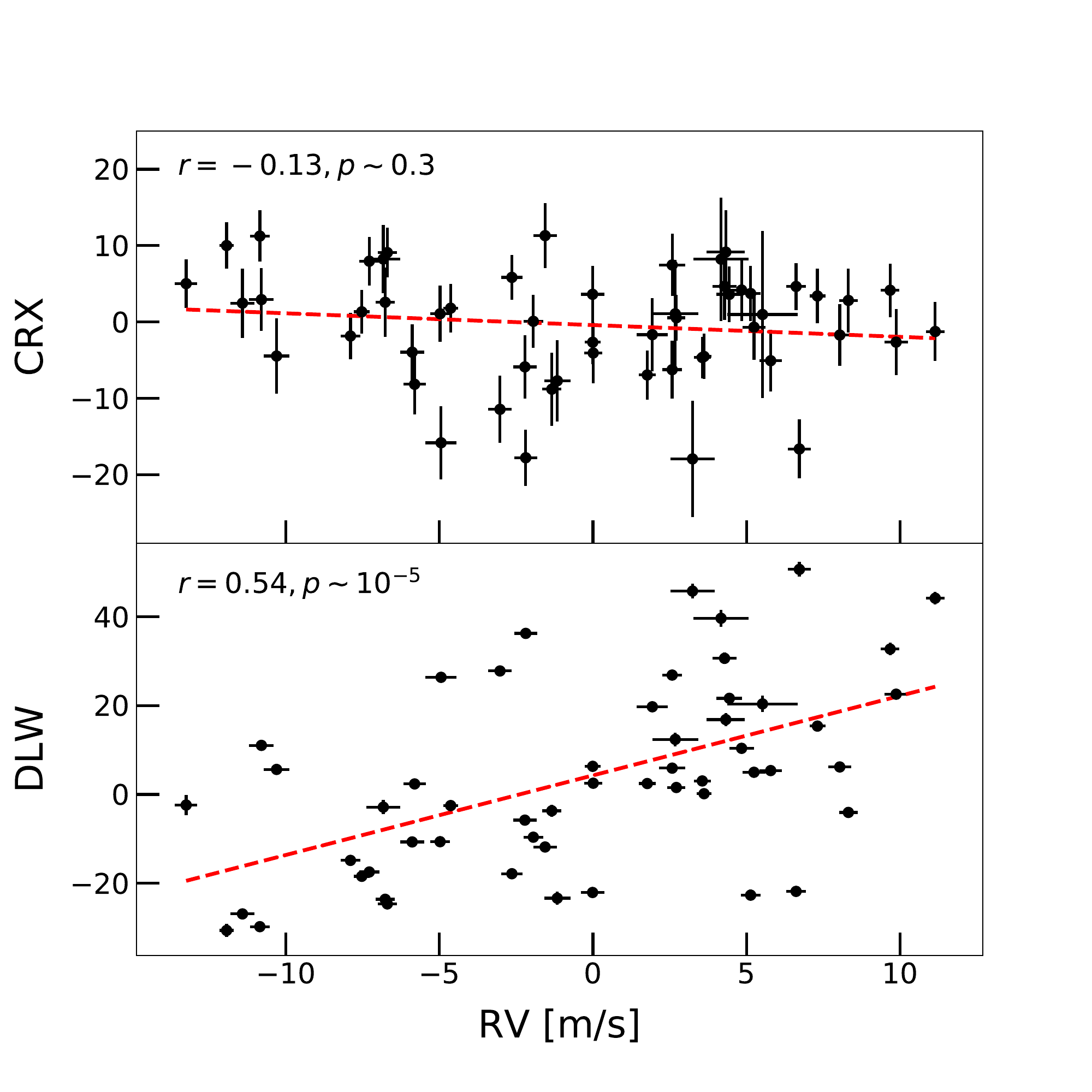}
\caption{Chromatic RV index (CRX) and differential line width (DLW) as a function of HARPS RV with the Pearson's correlation indexes and p-values shown on the upper left. The correlation fits are shown as red dashed lines. The correlation between DLW and RV indicates that the stellar activity has an effect on the HARPS Doppler signals (see Section \ref{rotation}).} 
\label{hai}
\end{figure}

\begin{figure}
\centering
\includegraphics[width=0.5\textwidth]{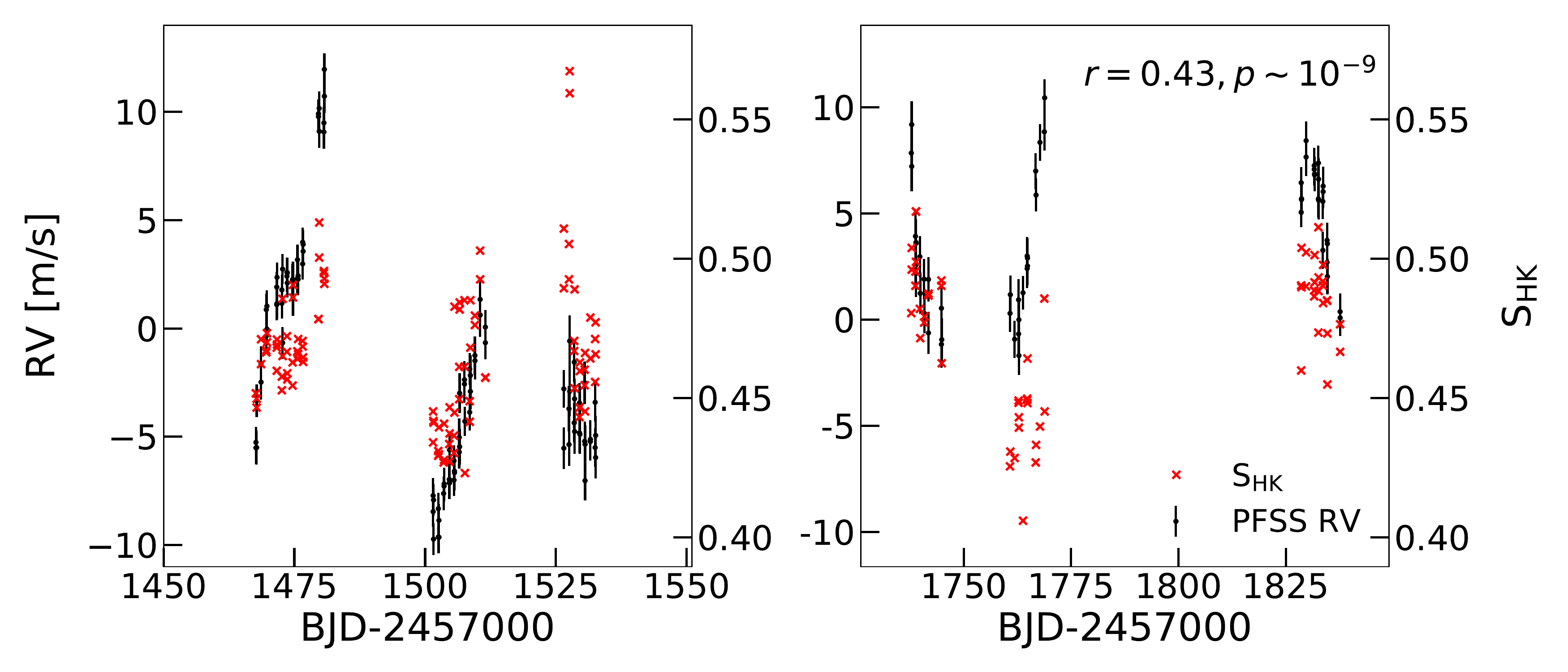}
\includegraphics[width=0.5\textwidth]{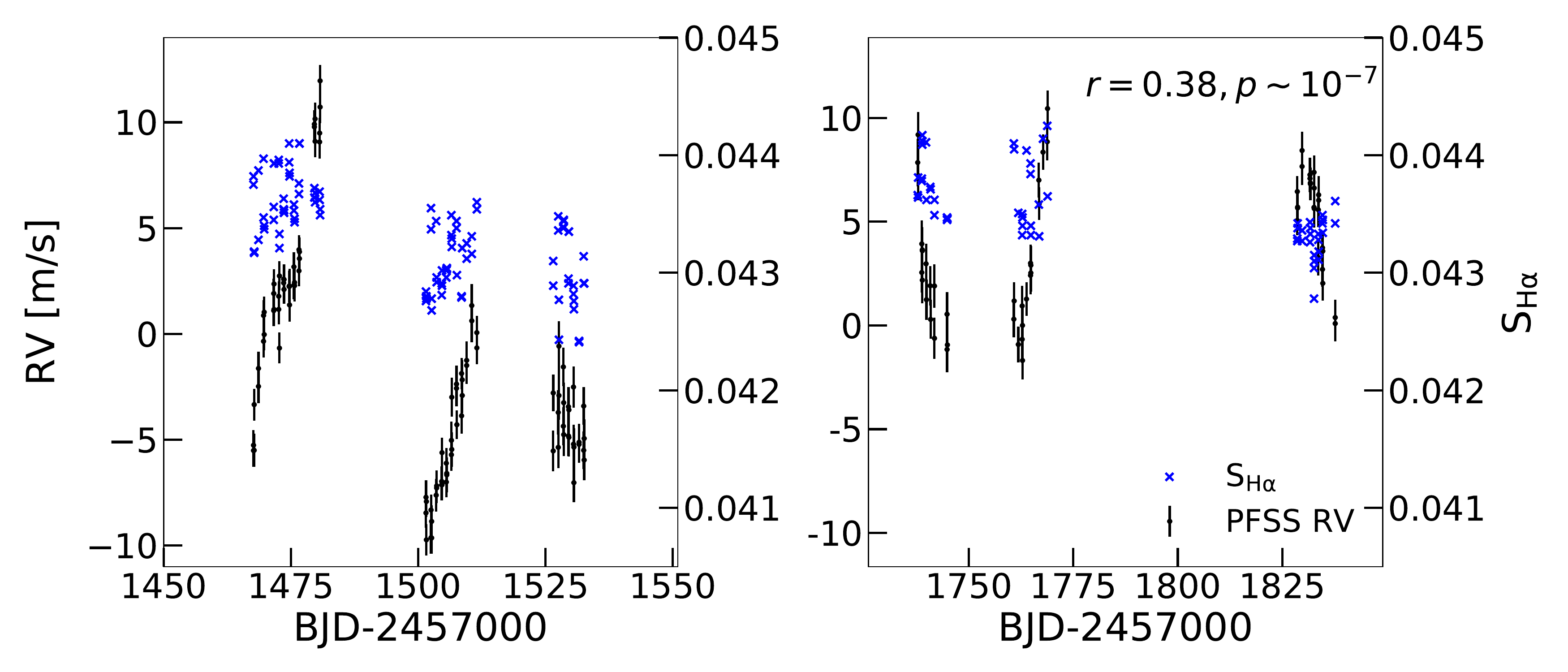}
\caption{{\it Top panels}: RV and stellar activity indicator $S_{\rm HK}$ versus time for PFSS only data. {\it Bottom panels}: RV and $S_{\rm H\alpha}$ versus time. The Pearson's correlation indexes and the corresponding p-values are shown on the upper right. The clear similar trend of RV and $S_{\rm HK}$, $S_{\rm H\alpha}$ makes the stellar rotation influence require particular care.} 
\label{SHK}
\end{figure}

In order to determine whether the stellar rotation has a significant effect on the Doppler signals, we measure the Pearson's {\it{r}} correlation coefficients between the temporal series of RV and CRX, DLW. Figure \ref{hai} shows the CRX and DLW as a function of RV. We find HARPS RVs have a weak anti-correlation\ ($r = -0.13$, p-value $\sim 0.3$) with CRX but a strong correlation\ ($r = 0.54$, p-value $\sim 10^{-5}$) with DLW. Clear correlations between RV and $S_{\rm HK}$, $S_{\rm H\alpha}$ are also found in the PFSS-only data\ ($r = 0.43$, p-value $\sim 10^{-9}$; $r = 0.38$, p-value $\sim 10^{-7}$), which suggests that we need to mitigate the stellar jitter caused by rotation (Figure \ref{SHK}).  Although D19 estimated that the stellar variability has a minor effect on the RV signals ($\sigma_{s,\rm RV} \approx 1.3$ m/s) based on the standard deviation of the \tess\ fluxes ($\sigma_{s, {\rm TESS}} \sim 0.0013$), our evaluation of the stellar activity of \tar\ above prompted us to reanalyze this system. We show that the stellar activity has a $\sigma_{pp,\rm RV}\approx6$ m/s contribution to the RV according to the peak-to-peak flux variation ($\sigma_{pp,{\rm TESS}} \sim 0.006$) in Section \ref{rv-only}.

\subsection{Fitting the Transits and the Stellar Activity in TESS Photometry}\label{photometry-only}

In order to constrain the stellar rotation period from photometrirc data directly, instead of fitting a basis spline to detrend the \tess\ light curve as in D19, we perform a joint fit of the light curve with a Gaussian Process\ (GP) model using {\bf juliet} \citep{juliet}. The transit model is created by {\bf batman} \citep{Kreidberg2015} while the GP is modelled using {\bf celerite} \citep{Foreman2017}. Dynamic nested sampling is performed using the public package {\bf dynesty} \citep{Higson2019,Speagle2019}.

Though \tess\ has high photometric precision, light contamination from nearby stars is of concern due to its large pixel scale (21$''$ per pixel, \citealt{Ricker2014}), which could pollute the aperture. To correct the flux dilution effect,\ {\bf juliet} enables one to fit a model with the form :
\begin{equation}
    \mathcal{M}(t)=\left[\mathcal{T}(t)D+(1-D)\right]\left(\frac{1}{1+D\cdot M} \right),
\end{equation}
where $\mathcal{M}(t)$ and $\mathcal{T}(t)$ represent the photometric and transit models while $D$ and $M$ are the dilution factor and mean offset out-of-transit flux, respectively. We search the nearby stars of \tar\ ($G_{\rm rp}=6.95$) within 1$'$ in the \gaia\ DR2 \citep{GaiaHR2018} to estimate the flux dilution factor $D_{\rm TESS}$ in the \tess\ photometry. A relatively bright giant star (\gaia\ DR2 4673947071237511552, $G_{\rm rp}=9.05$) 22$''$ away is expected to have a significant contribution to the contaminated flux while three faint neighbors located at $>35''$ with $G_{\rm rp}\gtrsim 16.0$ have essentially no measurable effect. As the \gaia\ $G_{\rm rp}$ band and the \tess\ band are similar, we simply evaluate the $D_{\rm TESS}\approx 0.87$ using Equation 6 in \cite{juliet}:
\begin{equation}
    D=\frac{1}{1+ \sum_{n} F_{n}/F_{T}},
\end{equation}
where $F_{T}$ and $F_{n}$ represent the flux of target and contamination sources. This value is consistent with 0.8605, the one reported in the \tess\ Input Catalog (TIC) V8 \citep{Stassun2017tic,Stassun2019tic}. Hence we set a Gaussian prior on $D_{\rm TESS}$ during the joint-fit based on the estimate above with the width $\sigma=0.01$. In addition, instead of directly fitting for the planet-to-star radius ratio ($R_{P}/R_{\star}$), orbital inclination ($i$) and scaled separation distance ($a/R_{\star}$), we sample points using the new parametrizations $r_{1}$ and $r_{2}$ to keep all values physically plausible \citep{Espinoza2018}. Because of the weak constraint on eccentricity from photometric data alone, we assume circular orbits for both \tar b and \can , place a non-informative log-uniform prior on stellar density and fix the orbital eccentricity to zero in our transit fit. We apply the quadratic limb-darkening law and sample the coefficients ($q_{1}$ and $q_{2}$) between 0 and 1 uniformly as described in \cite{Kipping2013}. We include a flux jitter term $\rm \sigma_{TESS}$ to describe the additional white noise in the \tess\ photometry. 

To model the photometric variations in the \tess\ light curve, we use GP regression with a rotation kernel formulated by \cite{Foreman2017}:
\begin{equation}
    k_{i,j}(\tau) = \frac{B}{2+C}e^{-\tau/L}\left[{\rm cos}\left(\frac{2\pi\tau}{P_{\rm rot}}\right) + (1+C) \right],
\end{equation}
where $B$ defines the GP covariance amplitude, $C$ is a balance parameter for the periodic and the non-periodic parts, $\tau=|t_{i}-t_{j}|$ is the time-lag between data point $i$ and $j$, and $L$ and $P_{\rm rot}$ represent the coherence timescale and the stellar rotational period, respectively. We adopt uninformative, wide log-uniform priors to the GP parameters except for the periodic timescale $P_{\rm rot}$, where we choose a relatively narrow uniform prior between 30 and 40 days according to our findings in Section~\ref{rotation}. We find a best fit of $P_{\rm rot}=31.2\pm1.5$ d and a radius of $2.86\pm0.20\ R_{\oplus}$ in the GP+transit joint fit, combined with the stellar radius $R_{\star}=0.695\pm0.030$ $R_{\odot}$ taken from D19. We note that D19 was published before TICv8 came out thus it used the original dilution value from TICv7. Our radius estimate here is larger than the radius of $2.61^{+0.17}_{-0.16}\ R_{\oplus}$ reported by D19 due to an updated dilution factor in this work (TICv8). In addition, combined with the stellar effective temperature $T_{\rm eff}=4640\pm100$ K retrieved from D19, assuming albedo is equal to 0, we find \tar b and \can\ have an equilibrium temperature of $407\pm22$ K and $676\pm36$ K, respectively.

\begin{figure*}
\centering
\includegraphics[width=\textwidth]{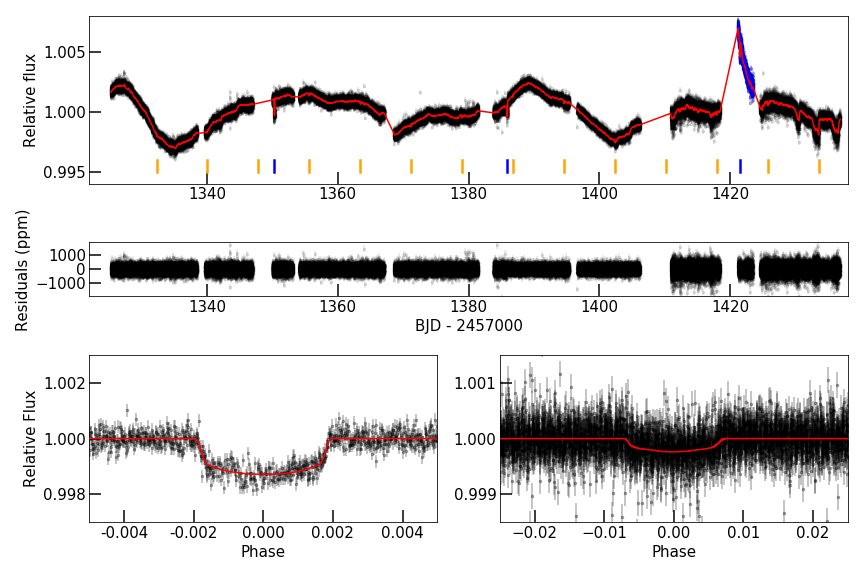}
\caption{{\it Top panel}: The full \tess\ discovery light curve (black points) of \tar\ along with the best fit GP+transit model (solid red line). Expected transits of \tar b and \can\ are marked in blue and orange ticks. The highlighted points between 1420 and 1440 have a strong downturn feature probably caused by systematics, although it did not affect the results of our GP model (see Section \ref{photometry-only}). {\it Middle panel}: Residuals after subtracting the best-fit model. {\it Bottom panels}: Phase-folded light curve of \tar b and \can, respectively, with best-fit GP model subtracted. The best-fit transit models are shown as red solid lines.} 
\label{transit+GP}
\end{figure*}

\begin{figure}
\centering
\includegraphics[width=0.5\textwidth]{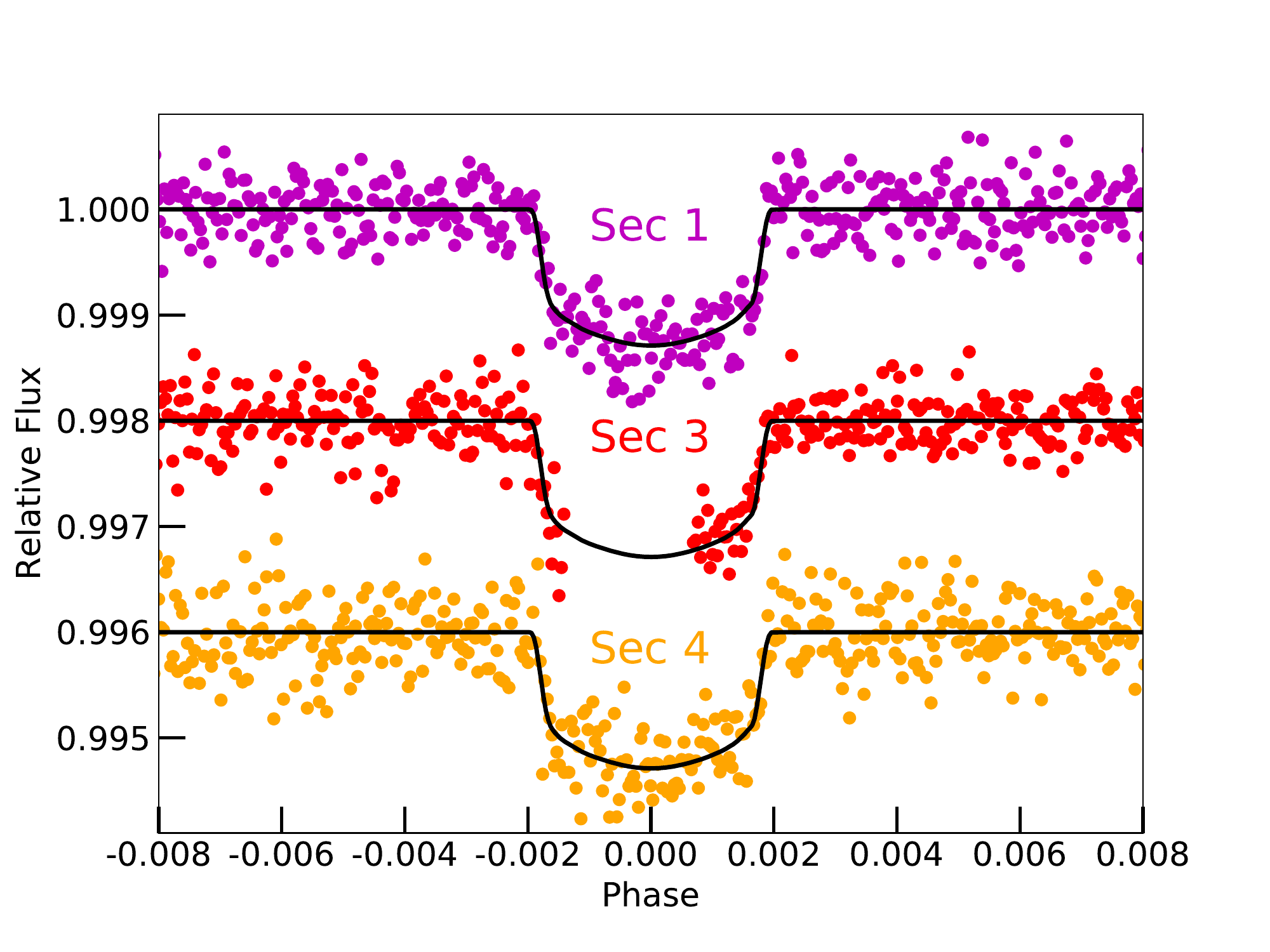}
\caption{The phase-folded \tess\ light curves of \tar\ from different sectors with best-fit transit models shown as black solid lines. The best-fit GP model has been subtracted. See Section \ref{photometry-only}.} 
\label{TESS_all_transit}
\end{figure}

As mentioned in Section~\ref{rotation}, the second segment in Sector 4 might be subject to strong systematic noise, and thus we exclude this feature so that it does not bias our estimate of $P_{\rm rot}$. Therefore, we fit a GP-only model to the \tess\ photometry after excluding that abnormal part and masking out all transits, and we obtain $P_{\rm rot}=31.7\pm3.4$ d, consistent with our current findings in the GP+transit joint fit. We list the best-fit parameters and the corresponding prior settings in Table \ref{trangppriors}. The \tess\ light curve of \tar\ along with the best-fit GP+transit model are shown in Figure \ref{transit+GP}. We present each transit of \tar b in Figure \ref{TESS_all_transit}. 

\begin{table*}
    \centering
    \caption{Model parameters of \tar b, \can\ and the best-fit values in the GP+transit joint fit analysis.}
    \begin{tabular}{lccr}
        \hline\hline
        Parameter       &Best-fit Value       &Prior     &Description\\\hline
        \it{Planetary parameters}\\
        $P_{b}$ (days)   &$35.6133^{+0.0005}_{-0.0006}$  
        &$\mathcal{N}^{[1]}$ ($35.6$\ ,\ $0.1^{2}$)
        &Orbital period of \tar b.\\
        $T_{0,b}$ (BJD)    &$2458350.3122^{+0.0007}_{-0.0007}$ 
        &$\mathcal{N}$ ($2458350.3$\ ,\ $0.1^{2}$) 
        &Mid-transit time of \tar b.\\
        $r_{1,b}$    &$0.7865^{+0.0310}_{-0.0340}$ 
        &$\mathcal{U}^{[2]}$ (0\ ,\ 1)
        &Parametrisation for {\it p} and {\it b}.\\
        $r_{2,b}$    &$0.0377^{+0.0011}_{-0.0010}$ 
        &$\mathcal{U}$ (0\ ,\ 1)
        &Parametrisation for {\it p} and {\it b}.\\
        $e_{b}$                     &0  &Fixed  &Orbital eccentricity of \tar b.\\
        $\omega_{b}$ (deg)          &90 &Fixed  &Argument of periapsis of \tar b.\\
        \\
        $P_{c}$ (days)   &$7.7902^{+0.0004}_{-0.0006}$  
        &$\mathcal{N}$ ($7.8$\ ,\ $0.1^{2}$)
        &Orbital period of \can.\\
        $T_{0,c}$ (BJD)    &$2458332.2783^{+0.0035}_{-0.0032}$ 
        &$\mathcal{N}$ ($2458332.3$\ ,\ $0.1^{2}$) 
        &Mid-transit time of \can.\\
        $r_{1,c}$    &$0.4780^{+0.1217}_{-0.0980}$ 
        &$\mathcal{U}$ (0\ ,\ 1)
        &Parametrisation for {\it p} and {\it b}.\\
        $r_{2,c}$    &$0.0149^{+0.0007}_{-0.0007}$ 
        &$\mathcal{U}$ (0\ ,\ 1)
        &Parametrisation for {\it p} and {\it b}.\\
        $e_{c}$                     &0  &Fixed  &Orbital eccentricity of \can.\\
        $\omega_{c}$ (deg)          &90 &Fixed  &Argument of periapsis of \can.\\
        \\
        \it{\tess\ photometry parameters}\\
        $D_{\rm TESS}$     &$0.868^{+0.009}_{-0.008}$ 
        &$\mathcal{N}$ ($0.87$\ ,\ $0.01^{2}$)      &TESS photometric dilution factor\\
        
        $M_{\rm TESS}$ (ppm)    &$245^{+1159}_{-1142}$
        &$\mathcal{N}$ (0\ ,\ $0.1^{2}$)      &Mean out-of-transit flux.\\
        
        $\sigma_{\rm TESS}$ (ppm) &$166^{+1}_{-2}$
        &$\mathcal{J}^{[3]}$ ($10^{-6}$\ ,\ $10^{6}$)      &TESS additive photometric jitter term.\\
        
        $\rm q_{1}$                &$0.32^{+0.12}_{-0.09}$       &$\mathcal{U}$ (0\ ,\ 1)  &Quadratic limb darkening coefficient.\\
        $\rm q_{2}$                &$0.66^{+0.22}_{-0.31}$       &$\mathcal{U}$ (0\ ,\ 1)  &Quadratic limb darkening coefficient.\\
        \\
        \it{Stellar parameters}\\
        ${\rho}_{\star}$ ($\rm kg\ m^{-3}$)   &$4066^{+742}_{-690}$
        &$\mathcal{J}$ ($100$\ ,\ $\rm 100^{2}$) &Stellar density.\\
        \\
        \it{GP hyperparameters}\\
        $B_{\rm TESS}$\ (ppm)  &$3.8^{+1.4}_{-1.3}$
        &$\mathcal{J}$ ($10^{-6}$\ ,\ $100$)      &Covariance amplitude of GP component.\\
        $C_{\rm TESS}\ [10^{-6}]$  &$416.2^{+2152.2}_{-407.5}$
        &$\mathcal{J}$ ($10^{-6}$\ ,\ $100$)      &Factor of GP component.\\
        $L$ (days)  &$57.8^{+23.2}_{-19.9}$
        &$\mathcal{J}$ ($0.01$\ ,\ $100$)      &Exponential
        evolutionary timescale of GP component.\\
        $P_{\rm rot}$ (days)  &$31.2^{+1.6}_{-0.8}$
        &$\mathcal{U}$ ($30$\ ,\ $40$)      &Periodic
        timescale of GP component.\\
        \\ \hline
        \it{Derived parameters} &{\bf \tar b} &{\bf \tar c} &\\
        $R_{P}/R_{\star}$    &$0.0377^{+0.0011}_{-0.0010}$ &$0.0149^{+0.0008}_{-0.0007}$ &Planet radius in units of stellar radii.\\
        ${b}$    &$0.676^{+0.046}_{-0.051}$ &$0.238^{+0.190}_{-0.149}$ &Impact Parameter.\\
        $a/R_{\star}$    &$64.84^{+3.73}_{-3.89}$  &$23.54^{+1.35}_{-1.41}$ &Semi-major axis in units of stellar radii.\\
        $i$ (deg)    &$89.40^{+0.07}_{-0.08}$ &$89.44^{+0.36}_{-0.52}$ &Inclination angle.\\
        $a$ (AU) &$0.209^{+0.022}_{-0.021}$ &$0.076^{+0.008}_{-0.008}$ &Semi-major axis.\\
        $R_{P}$ ($R_{\oplus}$) &$2.86^{+0.21}_{-0.20}$ &$1.13^{+0.11}_{-0.10}$  &Radius.\\
        $T_{\rm eq}$ (K)    &$407^{+22}_{-19}$ &$676^{+36}_{-32}$ &Equilibrium temperature.\\
        \hline\hline 
    \end{tabular}
    \begin{tablenotes}
    \item[1]  [1]\ $\mathcal{N}$($\mu\ ,\ \sigma^{2}$) means a normal prior with mean $\mu$ and standard deviation $\sigma$. 
    \item[2]  [2]\ $\mathcal{U}$(a\ , \ b) stands for a uniform prior ranging from a to b.
    \item[3]  [3]\ $\mathcal{J}$(a\ , \ b) stands for a Jeffrey's prior ranging from a to b.
    \end{tablenotes}
    \label{trangppriors}
\end{table*}

\subsection{RV Modeling}\label{rv-only}
\par We choose to fit the RVs independently of the transit fit with priors informed by the photometric analysis. To fit the RV measurements, {\bf juliet} utilizes the {\bf radvel} algorithm to create the Keplerian orbit model \citep{Fulton2018}. We compare different RV models based on the Bayesian model log-evidence ($\ln Z$) calculated by the {\bf dynesty} package. Overall, we consider a model to be favoured if $\Delta \ln Z>2$, and strongly supported if $\Delta \ln Z$ is greater than 5 \citep{Trotta2008,Luque2019}. 

\subsubsection{1pl model for all data}
\par First, the RV perturbation caused by \can\ is probably unmeasurable given the RV precision of our dataset. We estimate the planet mass to be $1.5\pm0.5\ M_{\oplus}$ based on the empirical mass-radius relation from \cite{Chen2017}, which corresponds to a radial velocity semi-amplitude $K_{c}$ of $\sim$ 0.6 m/s, assuming a circular orbit. This expected signal is beyond the detection limit of HARPS and PFS. Hence we choose to exclude \can\ in our RV analysis and start with a one planet (i.e., \tar b; 1pl for short) model fit (denoted as M1). For each instrument, we model the RV offset and the RV jitter term accounting for additional white noise independently. We fix the RV slope and quadratic terms\footnote{We fix the RV slope and quadratic terms to zero because our model comparison using Bayesian evidence suggests that a model without such terms is strongly favored ($\Delta \ln Z = 13$).} equal to 0 and free $e\sin \omega$ and $e\cos \omega$ in all our runs. In this fit with a pure Keplerian model, we obtain $K_{b}=6.11\pm0.37$ m/s, which agrees with the $5.51^{+0.41}_{-0.33}$ m/s value reported by D19 within 2$\sigma$. We note here that the error bar does not decrease considerably after adding 147 more PFSS RVs in this simple Keplerian fit, as one would expect in the case of white noise. This is probably because the newly added PFSS RVs contain highly correlated noise due to the existence of stellar activity sampled at high cadence.

\subsubsection{GP+1pl model for all data}
However, as shown in Section~\ref{rotation}, we find a clear correlation between the RV data points and the stellar activity indicators. This indicates that the RVs may have been influenced by stellar activity thus biasing the Keplerian fit. To estimate the RV variation induced by the stellar activity, we employ Equation 2 in \cite{Vanderburg2016}:
\begin{equation}
    RV_{pp} \simeq F_{pp}\times v \sin i,
\end{equation}
where $RV_{pp}$ and $F_{pp}$ are the peak-to-peak variation of RV and flux, respectively.\footnote{We note that here we apply the original peak-to-peak amplitude $F_{pp}$ of the \tess\ light curve rather than the standard deviation $\sigma_{pp}$ used in D19.} We estimate the $v \sin i$ to be $\sim 1.0$ km/s based on the stellar radius and derived rotation period in Section \ref{rotation} (again, consistent with the $v \sin i$ estimated from the HARPS spectra; \citealt{Dragomir2019}). The \tess\ light curve gives a $F_{pp}$ $\sim 6$ ppt (parts per thousand), which translates into a $\sim 6$ m/s effect on the RVs. This estimated RV effect from the stellar activity is similar to the RV semi-amplitude of the planet orbit. Therefore, in order to make a robust mass estimation, we include a GP model accounting for the stellar activity signals and perform a GP + one planet (GP+1pl; \tar b, M2) fit. 

\par Adopting the assumption that both the light curve and the RV curve should have the same frequency structure in the covariance function representing the stellar activity, we take the posteriors of coherence timescale $L$ and periodic timescale $P_{\rm rot}$ from the photometry-only fit into consideration and set Gaussian priors on both parameters for the RV fit accordingly. We set uninformative, wide Jeffrey's priors on the other GP parameters, similar to our transit fit (see Table~\ref{rvgppriors_GP_1pl}). We do not take the $\sigma_{P_{\rm rot}}=1.5$~d directly from the transit fit posterior but instead choose to set a relatively wide prior with $\sigma_{P_{\rm rot}}=4$~d. This choice of a relatively wide prior is to account for any potential systematic bias in the measurement of $P_{\rm rot}$ due to the relatively short baseline of the \tess\ photometry, as well as possible changes in $P_{\rm rot}$ due to changes in spot latitudes and differential rotation (see Section \ref{rotation}). Considering that the stellar rotation period is close to the orbital period of \tar b, which may present additional challenges in modelling, we fix the orbital period and the time of inferior conjunction of \tar b based on the transit ephemeris. We compare the results from global and instrument-by-instrument GP model and find $\Delta \ln Z\sim2$, and thus we employ the simpler global GP model with fewer degrees of freedom. 

\begin{figure*}
\centering
\includegraphics[width=\textwidth]{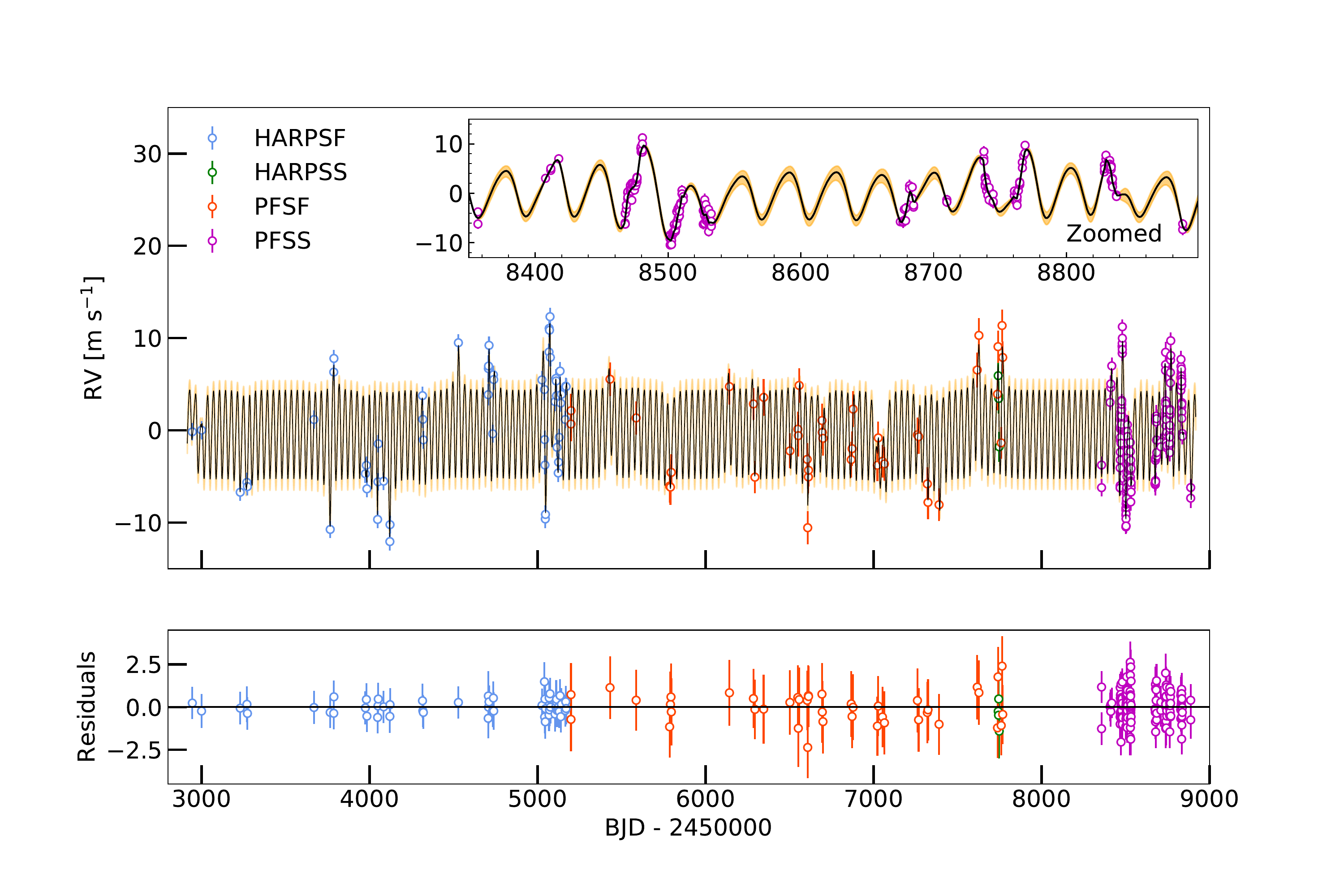}
\caption{{\it Top panel}: Pre-upgrade HARPS (HARPSF), post-upgrade HARPS (HARPSS), pre-upgrade PFS (PFSF) and post-upgrade PFS (PFSS) radial velocity data along with the best-fit GP+1pl model shown as a black solid line. Zoomed-in PFSS data are shown on the top right. {\it Bottom panels}: Residuals in m/s after subtracting the best-fit model. The error bars are the quadrature sum of the best-fit instrument jitter term and the measurement uncertainties for all RVs. }
\label{total_rv+GP_1pl}
\end{figure*}

\begin{figure}
\centering
\includegraphics[width=0.49\textwidth]{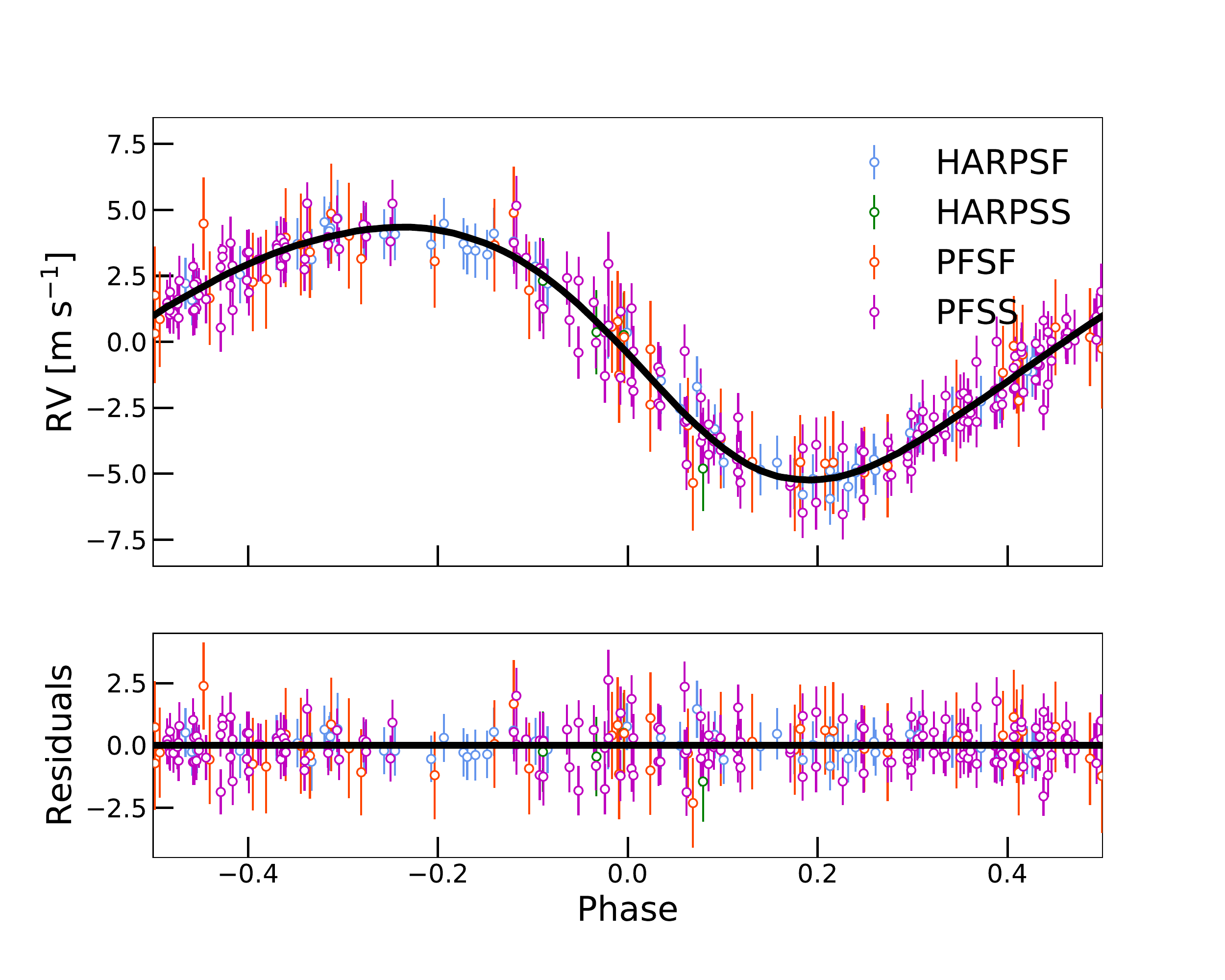}
\caption{RVs phase-folded to the period of \tar b (35.613 days). The RVs plotted here are the original RVs subtracting the estimated stellar activity signals represented by the best-fit GP model (see Section~\ref{rv-only}). The best-fit Keplerian model is shown as a black solid line. The residuals in m/s are shown in the lower panel.} 
\label{phase_rv+GP_1pl}
\end{figure}

\begin{figure*}
\centering
\includegraphics[width=\textwidth]{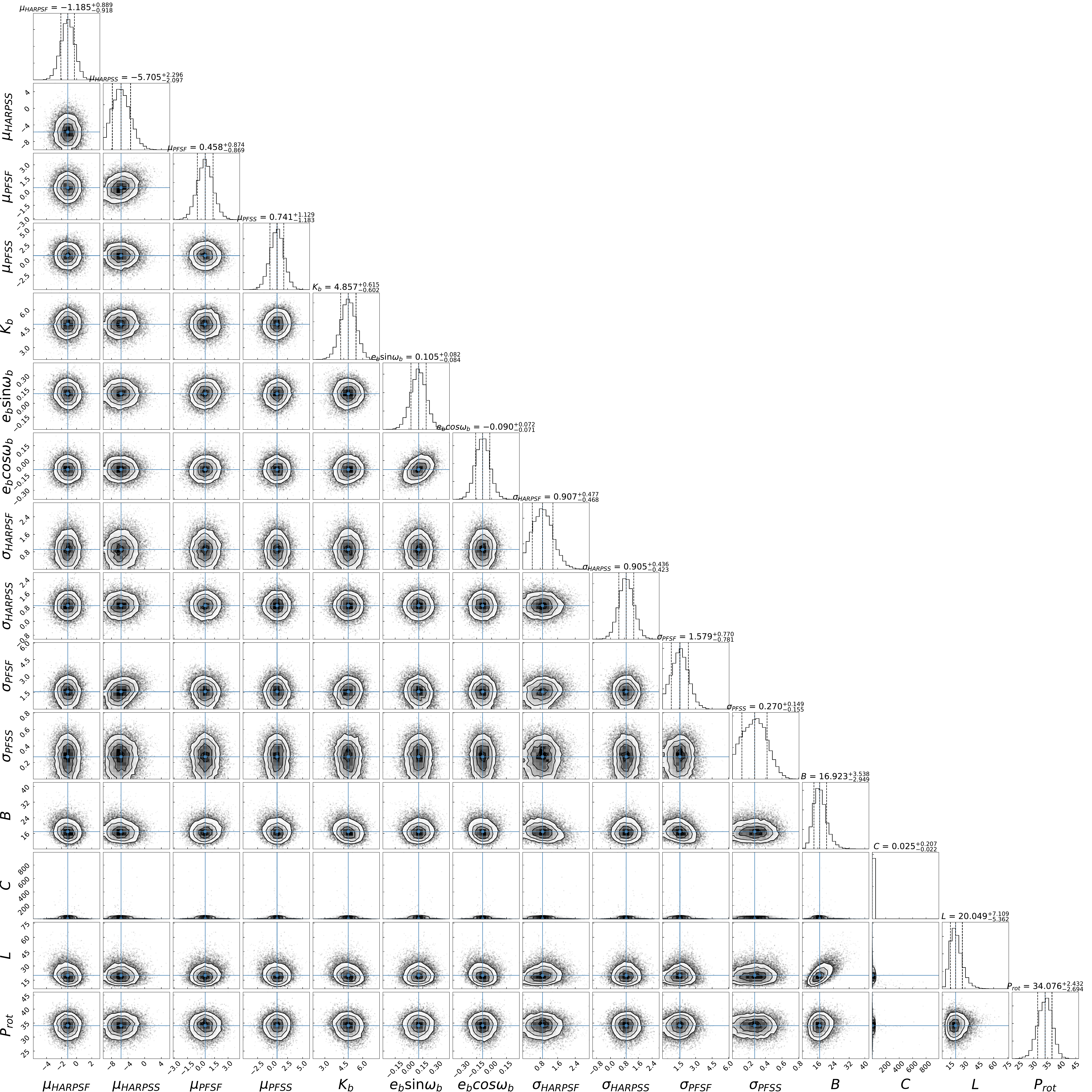}
\caption{Posterior distributions of the parameters from the fit with the GP+1pl model using all RV data.}
\label{posterior}
\end{figure*}

The RV residuals to the Keplerian-only model have prominent structures in the high-cadence PFSS data, suggesting strong correlated noise. After adding the GP model, such structures no longer exist. Compared with the previous Keplerian-only 1pl model, the GP+1pl model has a significant improvement in the Bayesian model log-evidence ($\Delta \ln Z=\ln Z_{\rm GP+1pl}-\ln Z_{\rm 1pl}=181$). Therefore, we consider the GP+1pl model (M2) to be favoured over the Keplerian 1pl model (M1). This best-fit GP+1pl model has an RV semi-amplitude $K_{b}$ of $4.86\pm0.61$ m/s and an eccentricity of $0.164\pm0.062$, leading to a mass measurement of $20.0\pm2.7\ M_{\oplus}$ (assuming the host star has a mass of M$_{\star}=0.73\pm0.07$ M$_{\odot}$; D19). This value is a little smaller than but still consistent with the previous estimate $22.7^{+2.2}_{-1.9}\ M_{\oplus}$ in D19. The best-fit RV model and the corresponding phase-folded RV curve for \tar b are shown in Figures \ref{total_rv+GP_1pl} and \ref{phase_rv+GP_1pl}. Figure \ref{posterior} shows the posterior distributions of the parameters. We report all relevant $\ln Z$ of different models in Table \ref{modelcomp} and all the best-fit parameters and their prior settings for this GP+1pl model in Table~\ref{rvgppriors_GP_1pl}. A recent work from \cite{Kosiarek2020} suggests exercising caution when using periodograms or ACF method to determine a stellar rotation period as it was not always the highest peak in these plots. Since the \tess\ photometry only covers $\sim 3$ stellar rotation cycles of \tar, we perform additional tests to see whether different prior settings on $P_{\rm rot}$ would have a significant effect on the measurement of $K_{b}$. We run two more GP+1pl fits for all RV data with 1) a much narrower Gaussian prior on $P_{\rm rot}$ centering at 31~d with $\sigma_{P_{\rm rot}}=1$~d; 2) a wide uniform prior between 25 and 60 d. We obtain $K_{b}=4.85\pm0.63$~m/s and $K_{b}=4.89\pm0.66$~m/s, respectively, which agree with the previous value derived from the original prior setting within $1\sigma$, indicating that $K_{b}$ is not sensitive to our prior choice on $P_{\rm rot}$. 

\subsubsection{GP+2pl model for all data}
\par We then run a GP+2pl fit to put a mass constraint on \can. We fix the orbital periods of both planets (\tar b and \can), set a narrow uniform prior on $K_{c}$ between 0 and 3 m/s and retain the same priors on the GP hyperparameters with the GP+1pl model. We fit a circular orbit for \can\ and obtain $K_{c}=0.65\pm0.26$\ m/s, which corresponds to a 3$\sigma$ mass upper limit of $\sim 3.5\ M_{\oplus}$ estimated using the 99.7\% value of $K$ in the posterior distribution.

\subsubsection{GP+1pl model for PFSS-only data}
\par
\begin{figure*}
\centering
\includegraphics[width=\textwidth]{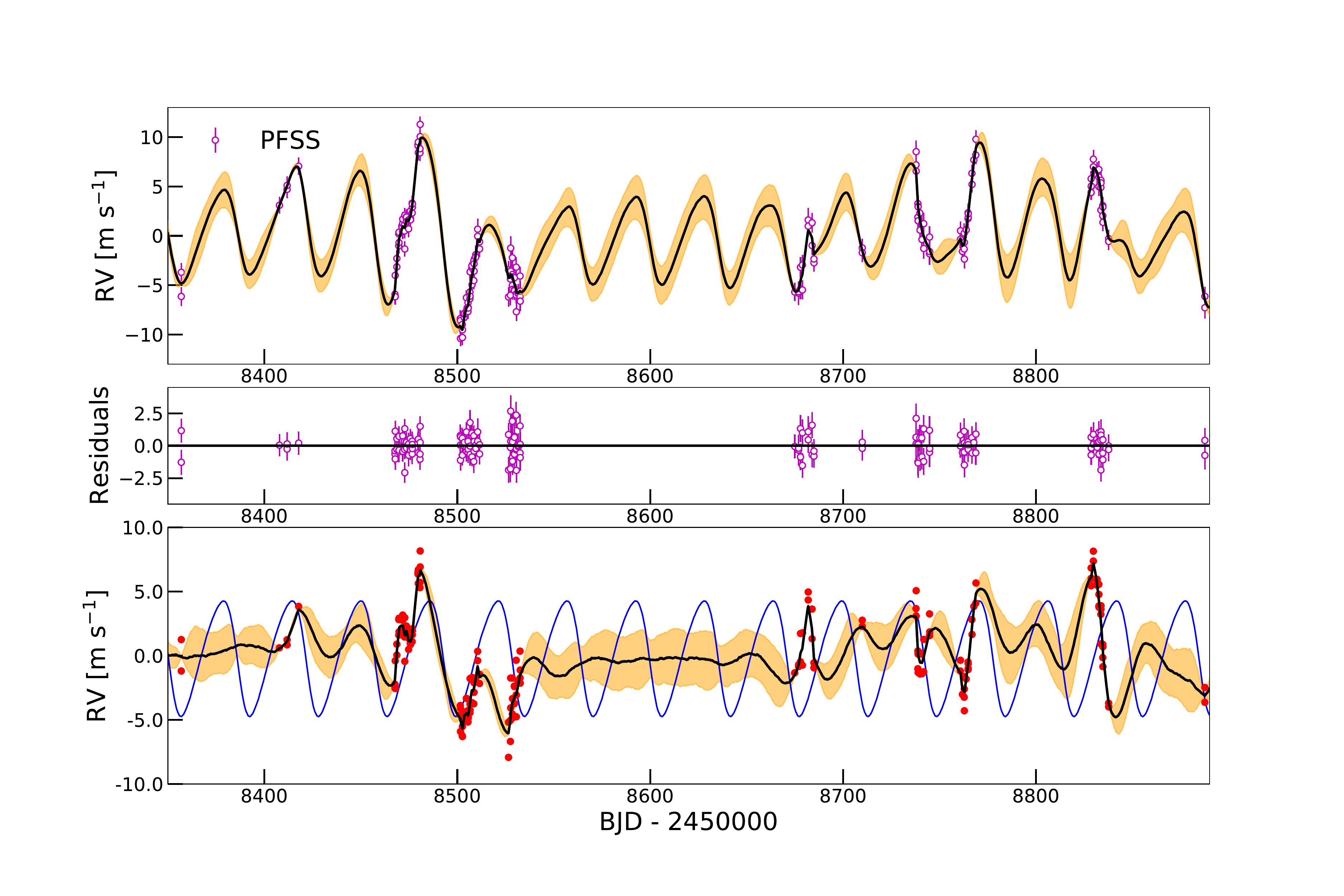}
\caption{{\it Top panel}: PFSS data along with the best-fit GP+1pl model. The shaded yellow region represents the 1$\sigma$ estimate. {\it Middle panel}: RV residuals after subtracting the best-fit model. {\it Bottom panel}: The red points are the total RV subtracting the best-fit Keplerian component of the GP+1pl fit for PFSS-only data. The black line is the predicted GP curve with a semi-amplitude $\sim 6$ m/s. The best-fit Keplerian orbit is shown as the blue curve.} 
\label{PFSS_only}
\end{figure*}

 Finally, we repeat the GP+1pl fit for the PFSS data independently because: 1) the timescale of the stellar activity is much smaller than the total RV baseline ($>$ 6000 d); 2) only the PFSS data have a sufficiently high cadence that well samples the stellar activity signals. We obtain $P_{\rm rot}=32.4\pm3.6$~d which is consistent with our findings in Section \ref{rotation} and Section \ref{photometry-only}. The best-fit value of $K$ in this case, $K_{b}=4.9\pm1.2$\ m/s, agrees with our estimates above using the full dataset within 1$\sigma$. Figure \ref{PFSS_only} shows the total RV curve and the corresponding predicted GP model for this PFSS-only fit. The $\sim 6$ m/s semi-amplitude of the GP model suggests that the stellar rotation indeed has a significant contribution to the RVs. Although having similar periods, the planetary signal from \tar b and the best-fit GP model for stellar activity are not always in phase, which is expected given the stochastic nature of the activity signal caused by stellar rotation. This disentanglement in phase space is essentially what enables us to tease out the stellar activity signals and obtain a more robust estimate of the planet's orbit. We list the free parameters and their priors along with the best-fit values for this PFSS-only fit in Table \ref{rvgppriors_PFSS}. 

\par We also explore modeling options other than a GP to mitigate the effects of stellar activity. Given that the activity indicator $S_{\rm HK}$ has a strong correlation with the RVs in the PFSS-only data, we first decide to perform an independent joint fit along with linear activity model. The large scatter and the clear correlated noise in the RV residuals motivate us to use a quadratic activity model instead with the form:
\begin{equation}
    RV_{\rm tot}(t_{i})=\gamma + RV_{\rm Kep}(t_{i})+\alpha\cdot{S_{\rm HK}}(t_{i})+\beta\cdot{S_{\rm HK}^{2}}(t_{i}),
\end{equation}
where $\gamma$ is the RV zero point. $RV_{\rm Kep}$ represents the Keplerian signal while $\alpha$, $\beta$ are the coefficients of the activity model. We apply the same RV prior settings as the ones that used in the previous GP+1pl fit for PFSS data. A Markov Chain Monte Carlo (MCMC) analysis is carried out to explore the posterior probability distribution of the orbital parameters with {\bf emcee} \citep{Foreman2013}. We initialize 250 walkers and run for 20,000 steps with the first 5000 steps abandoned as burnt-in sample. However, we do not find a clear improvement of the final result, which indicates that the current simple polynomial activity model can not successfully characterise the stellar rotation signal. 

\begin{table}
    \centering
    \caption{Model comparison of RV fits with juliet.}
    \begin{tabular}{ccccc}
        \hline\hline
        Model   &Type       &$P_{\rm planet}$ prior   &$\ln Z$    &$\Delta \ln Z$     \\\hline
        $\rm M1$    &1pl         &$\mathcal{F}^{[1]}_{\rm b}$ ($35.613$)             &-792.1 &180.8\\
        $\bf {M2}$    &{\bf GP+1pl}      &$\mathcal{F}_{\rm b}$ ($35.613$)    &{\bf -611.3}   &{\bf 0}\\
        $\rm M3^{[2]}$    &2pl         &$\mathcal{F}_{\rm b}$ ($35.613$)             &-787.6  &176.3\\
        & &$\mathcal{J}_{\rm d}$ ($10^{2}$\ ,\ $\rm 10^{4}$)   &  &\\
        $\rm M4$    &GP+2pl      &$\mathcal{F}_{\rm b}$ ($35.613$)  &-610.3 &-1.0\\
        & &$\mathcal{U}_{\rm d}$ ($2000$\ ,\ $\rm 4500$)   &  & \\
        \hline\hline 
    \end{tabular}
    \begin{tablenotes}
    \item[1]  [1]\ $\mathcal{F}$($\mu$) means fixing the orbital period equal to $\mu$. 
    \item[2]  [2]\ See Section \ref{add_pl} for M3 and M4. 
    \end{tablenotes}
    \label{modelcomp}
\end{table}

\begin{table*}
    \centering
    \caption{Model parameters of \tar b and the best-fit values in the GP+Keplerian (GP + 1pl) analysis for all RV data.}
    \begin{tabular}{lccr}
        \hline\hline
        Parameter       &Best-fit Value       &Prior     &Description\\\hline
        \it{Planetary parameters}\\
        $\rm P_{b}$ (days)   &$35.613$
        &Fixed
        &Orbital period of \tar b.\\
        $\rm T_{0,b}$ (BJD)    &$2458350.312$ 
        &Fixed
        &Mid-transit time of \tar b.\\
        $e_{b}\sin \omega_{b}$ &$0.105^{+0.082}_{-0.084}$  &$\mathcal{U}$ ($-1$\ ,\ $1$)  &Parametrisation for $e$ and $\omega$ of \tar b.\\
        $e_{b}\cos \omega_{b}$  &$-0.090^{+0.072}_{-0.071}$ &$\mathcal{U}$ ($-1$\ ,\ $1$)  &Parametrisation for $e$ and $\omega$ of \tar b.\\
        \\
        \it{RV parameters}\\
        $\rm \mu_{HARPSF}$ ($\rm m\ s^{-1}$)    &$-1.18^{+0.89}_{-0.92}$ 
        &$\mathcal{U}$ ($-10$\ ,\ $10$)
        &RV offset for pre-upgrade HARPS (HARPSF).\\
        $\rm \sigma_{HARPSF}$ ($\rm m\ s^{-1}$)    &$0.91^{+0.47}_{-0.46}$ 
        &$\mathcal{U}$ ($0$\ ,\ $10$)
        &Jitter term for pre-upgrade HARPS (HARPSF).\\
        $\rm \mu_{HARPSS}$ ($\rm m\ s^{-1}$)    &$-5.70^{+2.30}_{-2.10}$ 
        &$\mathcal{U}$ ($-10$\ ,\ $10$)
        &RV offset for post-upgrade HARPS (HARPSS).\\
        $\rm \sigma_{HARPSS}$ ($\rm m\ s^{-1}$)    &$0.90^{+0.43}_{-0.42}$ 
        &$\mathcal{U}$ ($0$\ ,\ $10$)
        &Jitter term for post-upgrade HARPS (HARPSS).\\
        $\rm \mu_{PFSF}$ ($\rm m\ s^{-1}$)    &$0.46^{+0.87}_{-0.87}$ 
        &$\mathcal{U}$ ($-10$\ ,\ $10$)
        &RV offset for pre-upgrade PFS (PFSF).\\
        $\rm \sigma_{PFSF}$ ($\rm m\ s^{-1}$)    &$1.58^{+0.77}_{-0.78}$ 
        &$\mathcal{U}$ ($0$\ ,\ $10$)
        &Jitter term for pre-upgrade PFS (PFSF).\\
        $\rm \mu_{PFSS}$ ($\rm m\ s^{-1}$)    &$0.74^{+1.13}_{-1.18}$ 
        &$\mathcal{U}$ ($-10$\ ,\ $10$)
        &RV offset for post-upgrade PFS (PFSS).\\
        $\rm \sigma_{PFSS}$ ($\rm m\ s^{-1}$)    &$0.27^{+0.15}_{-0.15}$ 
        &$\mathcal{U}$ ($0$\ ,\ $10$)
        &Jitter term for post-upgrade PFS (PFSS).\\
        $K_{b}$ ($\rm m\ s^{-1}$)       &$4.86^{+0.61}_{-0.60}$
        &$\mathcal{N}$ ($5.5^{[1]}$\ ,\ $2^{2}$)
        &RV semi-amplitude of \tar b.\\
        
        \\
        \it{GP hyperparameters}\\
        $B$ ($\rm (m\ s^{-1})^{2}$)  &$16.923^{+3.539}_{-2.949}$
        &$\mathcal{J}$ ($10^{-3}$\ ,\ $10^{3}$)      &Covariance amplitude of the global GP component.\\
        $C$  &$0.025^{+0.207}_{-0.022}$
        &$\mathcal{J}$ ($10^{-3}$\ ,\ $10^{3}$)      &Factor of the global GP component.\\
        $L$ (days)  &$20.049^{+7.109}_{-5.362}$
        &$\mathcal{N}$ ($57.8$\ ,\ $50^{2}$)      &Exponential
        evolutionary timescale of the global GP component.\\
        $P_{\rm rot}$ (days)  &$34.1^{+2.4}_{-2.7}$
        &$\mathcal{N}$ ($31$\ ,\ $4^{2}$)      &Periodic
        timescale of the global GP component.\\
        
        \\ \hline
        \it{Derived parameters} &{\bf \tar b} &{\bf \tar c} &\\
        $e$ &$0.164^{+0.062}_{-0.058}$ &- &Orbital eccentricity.\\
        $\omega$ (deg) &$125^{+31}_{-30}$ &- &Argument of periapsis.\\
        $M_{P}$ ($M_{\oplus}$) &$20.0^{+2.7}_{-2.7}$ &$< 3.5^{[2]}$ &Mass.\\
        $\rho_{p}$ ($\rm g\ cm^{-3}$) &$4.7^{+1.9}_{-1.4}$  &$< 13.2$ &Density.\\
         \hline\hline 
    \end{tabular}
    \begin{tablenotes}
    \item[1]  [1]\ This initial guess value 5.5 m/s is taken from D19.
    \item[2]  [2]\ This is not a statistically significant measurement but the $3\sigma$ mass upper-limit.
    \end{tablenotes}
    \label{rvgppriors_GP_1pl}
\end{table*}

\section{Stellar Activity Simulation}\label{simulation}

As mentioned in Section \ref{rv-only}, the RV variation caused by the planet perturbation is at the same level as the stellar activity effect and the orbital period is similar to the rotation timescale. Such extreme entanglement calls for extra caution when determining how stellar activity affects our estimate of the Keplerian signal. With all of the above in mind, we perform simulations to evaluate the robustness of our GP+Keplerian model that disentangles these two different signals.

\par
\begin{figure}
\centering
\includegraphics[width=0.5\textwidth]{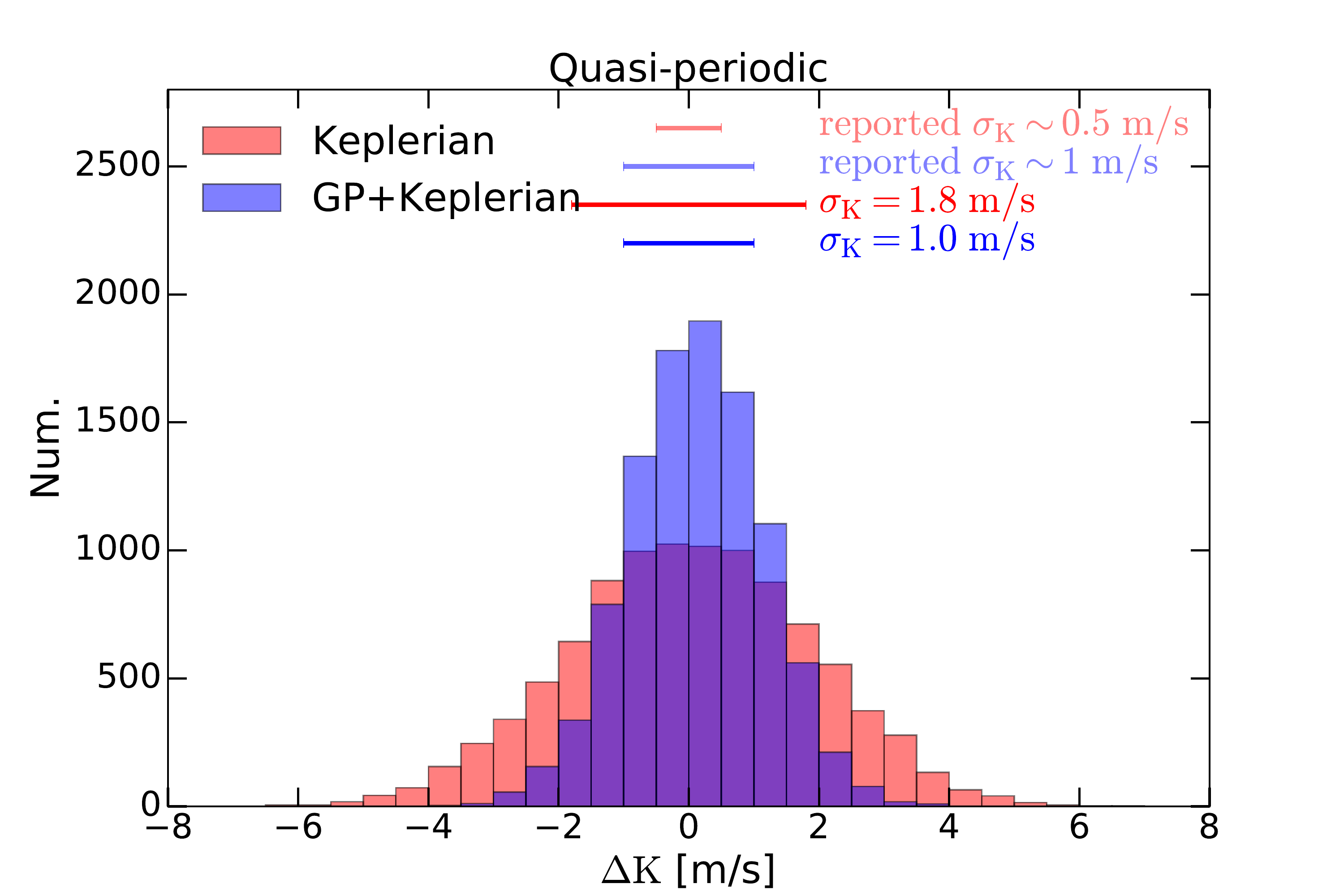}
\includegraphics[width=0.5\textwidth]{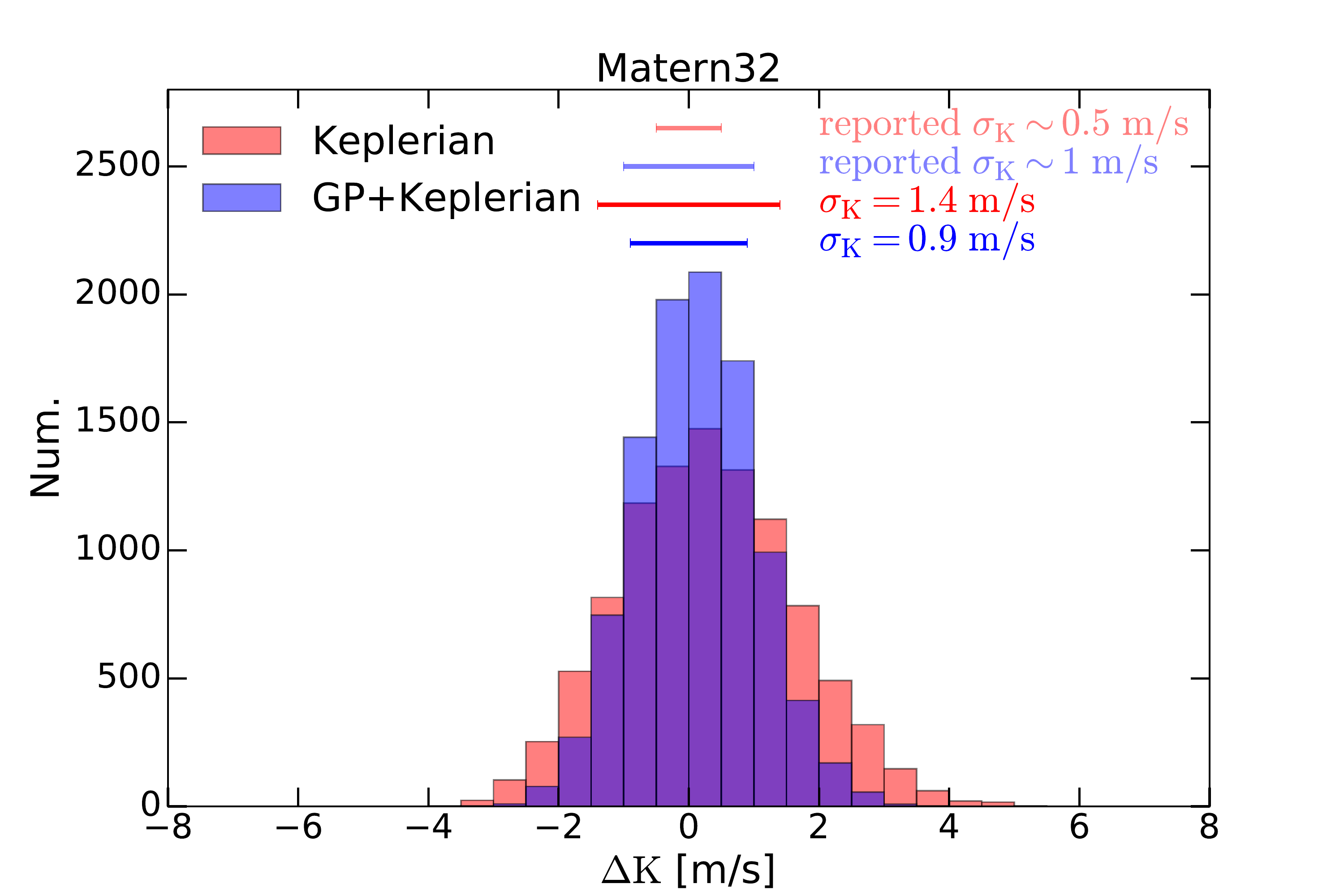}
\caption{K difference distribution of 10,000 simulated RV data derived from the 1pl and GP+1pl fit using the least-squares minimisation method. Different panels show the results with stellar activity signals generated by different GP kernels (see Section \ref{simulation}). The standard deviations of two $\Delta K$ distributions are shown as horizontal ticks above the histograms while the typical reported error bars from each kind of fit are the translucent ones (see Figure \ref{K_diff_nest}). The corresponding values are on the upper right.}
\label{K_diff_leastchi2}
\end{figure}

 We choose to perform the simulations in the context of the PFSS-only fit because of its high cadence sampling. We first create 10,000 sets of RV data from our best-fit GP models using {\bf celerite} based on the median GP hyperparameters values taken from the posterior distribution of our PFSS-only GP+1pl fit. After the superposition of a simulated Keplerian component representing \tar b with $K$ at 4.9 m/s\footnote{This value of $K$ is taken from the GP+1pl fit for PFSS data, which is consistent with what we find in the GP+1pl fit for all RV data.} and a random Gaussian white noise according to our best-fit RV jitter, we apply the GP+1pl fitting algorithm to these artificial data and minimize the likelihood function using the least squares method. For comparison, we repeat this with the Keplerian-only 1pl fit. The upper panel of Figure \ref{K_diff_leastchi2} shows the distribution of $\Delta K$, defined as the difference between the fitted $K$ and the input $K=4.9$~m/s in the simulation. For most cases of the GP+1pl model, we obtain $|\Delta K| \lesssim 1$ m/s with a standard deviation of $\sim$ 1.0 m/s. The evenly distributed $\Delta K$ near zero suggests that the GP+1pl model does not bias the real planet signal in our case. Without a GP model, we do not find a significant biased $K$ estimate but we do arrive at a slightly broader distribution with a standard deviation around 1.8 m/s, indicating a lower precision than the GP+1pl model. 

\begin{figure*}
\centering
\includegraphics[width=1\textwidth]{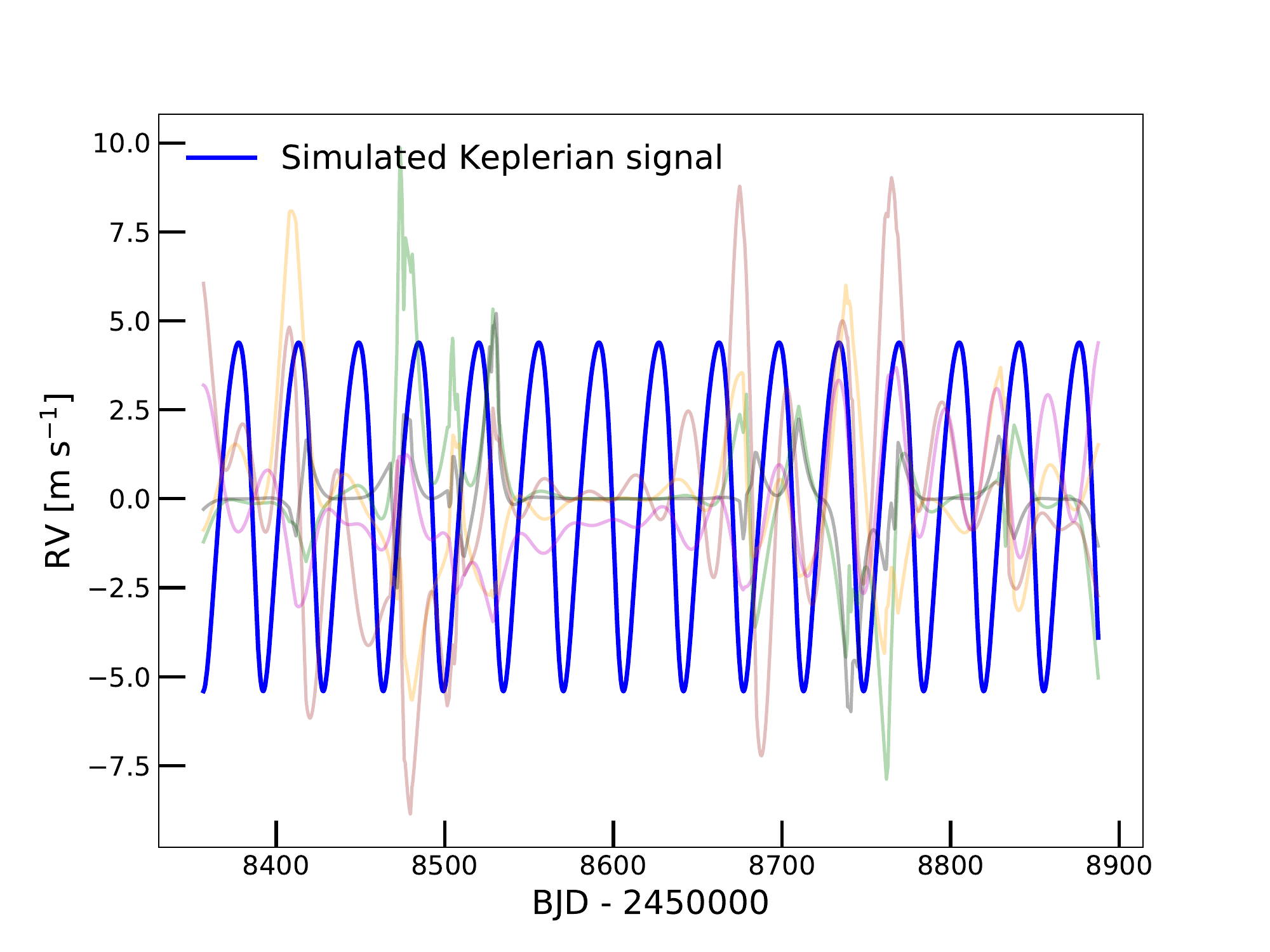}
\caption{Several examples of the injected GP models representing stellar activity signals from our simulations (various light colored curves) in comparison with the simulated planetary signal (the solid blue curve). See Section~\ref{simulation} for more detail.}
\label{simu_pred_GP}
\end{figure*}

We then generate another 10,000 sets of correlated noise regarded as stellar activity signals based on the {\bf george} package with a Matern32 kernel \citep{Foreman2015}:

\begin{equation}
    k_{i,j}(\tau) = A^{2}\left(1+\sqrt{3 \frac{\tau^{2}}{\lambda^{2}}}\right){\rm exp}\left(-\sqrt{3\frac{\tau^{2}}{\lambda^{2}}}\right),
\end{equation}
where $\tau$ is the time-lag, $A$ is the covariance amplitude and $\lambda$ is the correlation timescale. We fix $\lambda$ equal to the median value in the posterior distribution of $L$ from our PFSS-only GP+1pl fit as both $\lambda$ and $L$ contribute to the exponential term of two GP kernels. We randomly select several hyperparameter $A$ between 1 and 10, and examine the amplitude of the resulting signals. For convenience, we finally fix $A$ equal to 4, making the expected RV variation caused by the stellar activity close to the level we saw in the PFSS data (Figure \ref{PFSS_only}). Similarly, we add the same simulated Keplerian component and a random Gaussian white noise as mentioned above, and repeat the 1pl and GP+1pl fit. The corresponding $\Delta K$ distribution is shown in the lower panel of Figure \ref{K_diff_leastchi2}. By utilizing a different GP kernel to generate stellar activity signals, we notice that the standard deviation of $\Delta K$ derived from the 1pl fit has a small decrease but it is still much larger than the typical reported error bar ($\sim0.5$~m/s). Simultaneously, we obtain similar standard deviations of the $\Delta K$ distribution in the GP+1pl fit case, which are also consistent with the reported error bar ($\sim1$~m/s). We show the injected GP models of 5 simulated RV data sets in Figure \ref{simu_pred_GP}.

\begin{figure}
\centering
\includegraphics[width=0.5\textwidth]{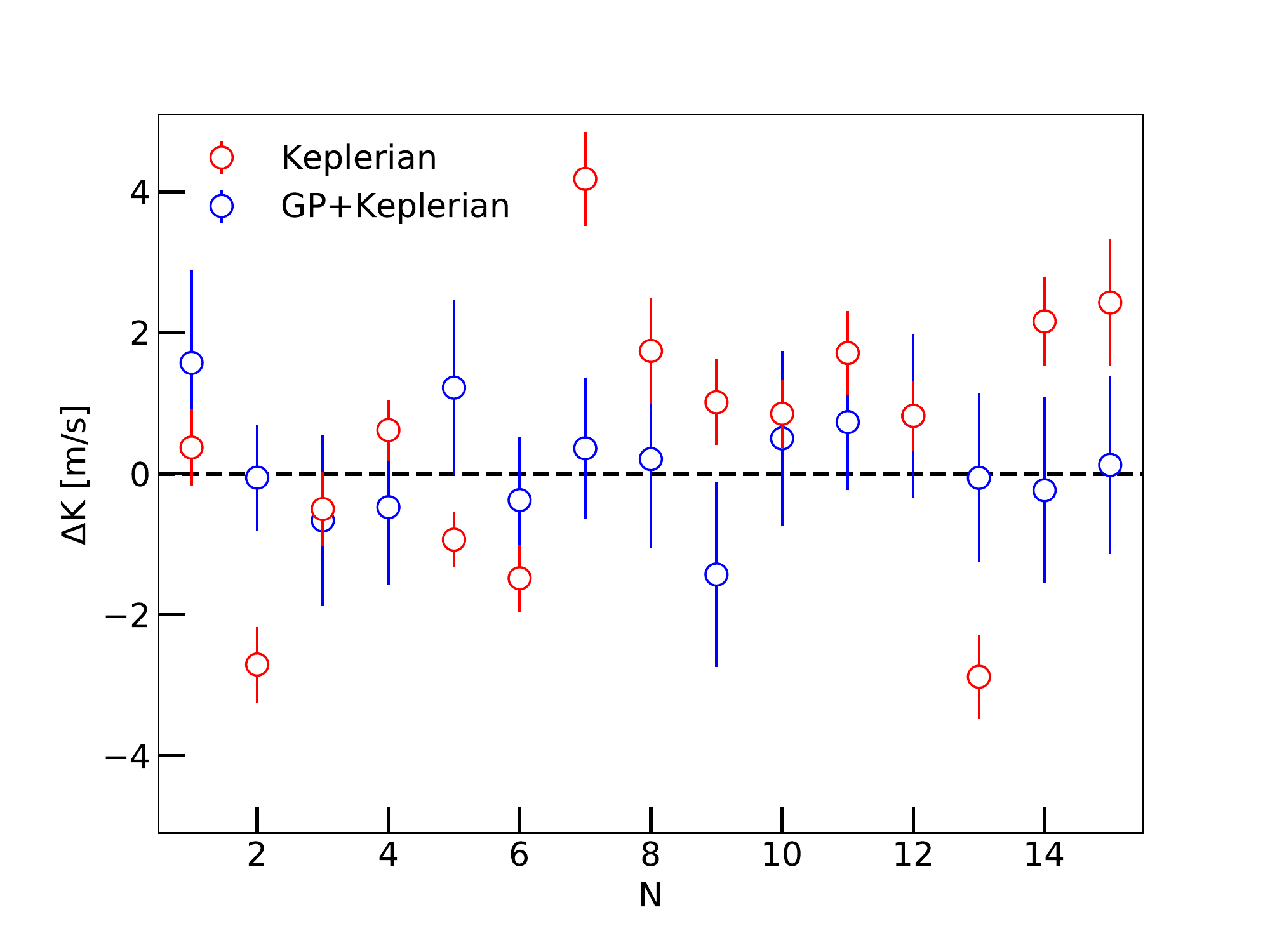}
\caption{Difference in $K$ between the best-fit and the input true value from the 15 independent simulated RV data sets for both GP+1pl and 1pl models. The 1$\sigma$ errors are derived from nested sampling. The large deviation and scatter of the results from the Keplerian-only model suggest that it is more uncertain and subject to an underestimation of error bars when measuring $K$.}
\label{K_diff_nest}
\end{figure}

Furthermore, we run 15 additional independent fits\footnote{We only run 15 simulations with nested sampling, because each run is relatively computational intensive.} with nested sampling for both GP+1pl and 1pl model to compare their accuracy and precision. The results are shown in Figure \ref{K_diff_nest}. Although the 1pl fit appears to give a more precise measurement of $K$ implied by a narrower posterior distribution with a typical error bar of $\sim 0.5$ m/s, the large deviation and scatter of the best-fit $K$ values from the input true value of $K$ (4.9 m/s) suggest unaccounted uncertainties in the modeling, which could lead to a biased estimate of the RV semi-amplitude with an underestimated error bar. Instead, in the GP+1pl fit, we find the estimated $K$ values are more often closer to the true value. In addition, the amount of scatter in $\Delta K$ is consistent with the reported scatter in the posterior distribution of $K$, at about 1~m/s. The result of these simulations suggest that the Keplerian-only model often reports an underestimated error bar for $K$, and it is also subject to more uncertainties when measuring $K$ than the model with stellar activity taken into account. 

\begin{figure}
\centering
\includegraphics[width=0.5\textwidth]{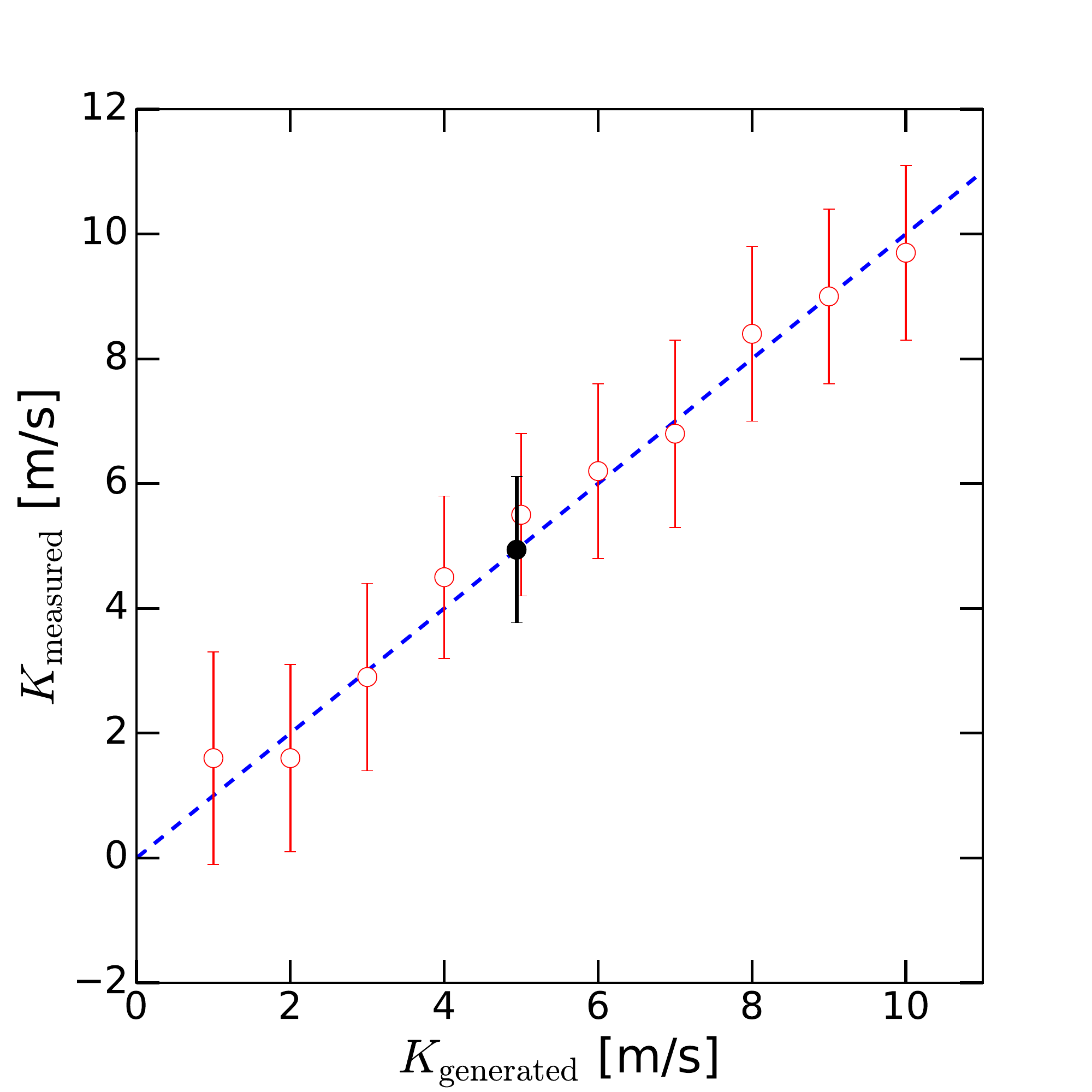}
\caption{The generated $K$ vs. measured $K$ using GP+1pl model fit. The 1$\sigma$ errors are derived from nested sampling. The black point represents our best-fit $K$ to the real PFSS data. The detection limit of our methodology is $2\sim4$ m/s, considering the $3\sigma$ detection. See Section~\ref{simulation} for more detail.}
\label{Kvsk}
\end{figure}

As a final test for the robustness of our model, we check the detection efficiency of our methodology. We fix the stellar rotation signal to our best-fit GP model to the real PFSS data and generate a total of 10 planet signals with various $K$ ranging from 1 to 10 m/s. Similarly, we add these two parts together along with a random Gaussian white noise for each sample. We perform the same GP+1pl fit for the artificial RV data with nested sampling\footnote{Note that we expand the prior bound on $K$ with a slightly larger $\sigma_{K}=5$ m/s.}. We present the final result in Figure \ref{Kvsk}, which suggests that the detection limit of our methodology is $2\sim4$ m/s as the size of the error bar makes the $3\sigma$ detection near 4 m/s. This detection limit is a lower level than the $\sim4.9$ m/s variation we found in the real RV data, validating the robustness of our measurement. We note that this added activity + white noise component is very similar to adding the residuals of the original RVs subtracting the best-fit Keplerian model, given that our GP model had a high goodness of fit to the activity component. The difference versus using the residuals is that we use a different set of additional Gaussian noise in each simulation at a different K value, randomizing the white noise component.

\section{Discussions}\label{discussion}

\subsection{\tar b as an Atmospheric Characterization Target}

Warm sub-Neptunes or super-Earths around nearby bright stars are extremely rare. According to data from the NASA Exoplanet Archive, among over 3000 confirmed exoplanets smaller than Neptune, only 27 are around stars with 2MASS $K_s$ band magnitude brighter than 10~mag. If we limit the equilibrium temperature of the planet to be lower than 500~K, then there are only 6 such planets, all around M dwarfs except for \tar b. \tar b also has the brightest host star in the $K_s$ band (5.4~mag) and by far the brightest in the optical $V$ band (8.1~mag), making it an attracting object for future atmospheric characterization.

\begin{figure}
\centering
\includegraphics[width=0.5\textwidth]{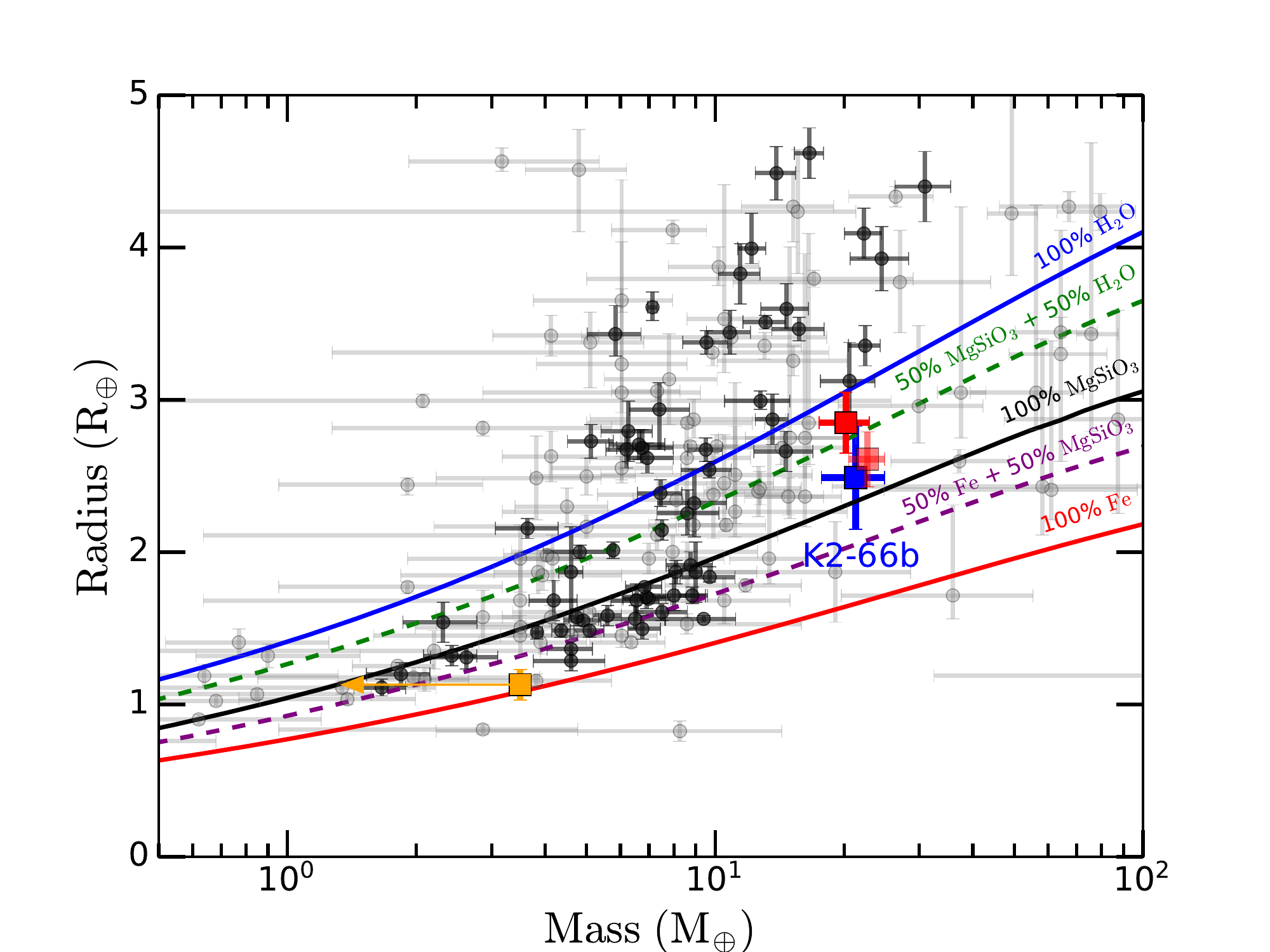}
\caption{Mass-radius diagram in Earth units. The updated position of \tar b is marked as a red square while the original position is shown as the translucent one based on D19. \can\ is shown as an orange square (3$\sigma$ mass upper limit). Black dots represent planets with precise measurement for mass and radius (uncertainty smaller than 20\% of the measured parameters) while gray points are planets with weak constraint on their radius or mass (data from NASA Exoplanet Archive; \citealt{Akeson2013}). The colored lines are taken from theoretical models for different planetary compositions in \citet{Zeng2013}.}
\label{mr}
\end{figure}

Having an accurate and precise measurement for the planetary mass is very important for reliably retrieving the atmospheric parameters, as demonstrated by the work of \cite{Batalha2019}. Our updated measurements for \tar b in this work provide the latest robust estimates for the density and surface gravity for this favorable target for atmospheric characterization. D19 reported \tar b\ as the second densest sub-Neptune with a density of $\rm 7.6^{+1.6}_{-1.3}\ g\ cm^{-3}$, after K2-66b \citep{Sinukoff2017}. However, we obtain a smaller mean density, $\rm 4.80^{+1.98}_{-1.40}\ g\ cm^{-3}$, combining our updated mass and radius estimates of \tar b. The estimated surface gravity of \tar b also decreased by $\sim 20\%$ with the updated mass and radius estimates. As shown in the mass-radius diagram (Figure \ref{mr}), \tar b is consistent with a rocky composition of magnesium silicate and water. We simply evaluate the radius of the rock core $R_{\rm core}=2.1\pm0.1\ R_{\oplus}$ according to Equation 1 in \cite{Lopez2014}, which implies that \tar b could hold a relatively thick gaseous envelope. 

\tar\ is a suitable source for ground based atmosphere characterization given its brightness. However, it is not an ideal target for \jwst. First, as noted by D19, \tar b does not have a very high Transmission Spectroscopic Metric\footnote{With the updated planetary parameters, we obtain $\rm TSM=75^{+46}_{-29}$ for \tar b.} (TSM; \citealt{Kempton2018}). Moreover, with a $J\sim6$ mag, HD 21749 sits on, or surpasses the edge of hard saturation for two of the Bright Object Time Series (BOTS) instruments of \jwst.

A \jwst\ integration is composed of a reset of the detector, followed by a sequence of non-destructively sampled \emph{groups}. The time it takes to read out one group, and reset the detector, is specified by each observing mode's unique frame time, $t_f$. Currently, the Astronomer's Proposal Tool\footnote{\url{https://www.stsci.edu/scientific-community/software/astronomers-proposal-tool-apt}} requires that each BOTS mode has at least two groups per integration. Therefore, the ``hard" saturation limit of each instrument is defined by whether the electron count of any pixel on the subarray has surpassed the full well limit at the end of two groups \citep{Pontoppidan2016,Batalha2017}. For NIRISS SOSS's smallest subarray (SUBSTRIP96, $t_f=2.2$~s), assuming an optimistic saturation limit of $100\%$ full well,  there would be 14,028 pixels saturated at the end of the first group. For NIRSpec G395H ($t_f=0.90$~s), there would be 1,265 pixels saturated at the end of the first group.  Given that G395H is NIRSpec's highest resolution and longest wavelength spectrograph, it would preclude observations from other NIRSpec BOTS modes (i.e. Prism, G140M/H, G235M/H, or G395M). Thus, even if the 2 group minimum requirement is decreased to 1 group, NIRISS and NIRspec observations would be unattainable. 

On the other hand, both NIRCam and MIRI would be free of saturation concerns. Using NIRCam's 4amp readout mode ($t_f=0.34$~s), a F322W2 observation could be made with 3 groups without surpassing the optimistic 100\% full well saturation limit. An observation with an integration time of $2$T$_{14}$, would result in a minimum precision of 67~ppm at 2.8 $\mu$m where F322W2's native resolution is R=1397 (or roughly 20~ppm if binned to R=100). Likewise, MIRI LRS ($t_f=0.159$~s) could also fit a 3 group observation assuming 100\% full well. Similarly, an observation with an integration time of $2$T$_{14}$ would result in a minimum precision of 23~ppm for MIRI LRS at its native resolution of R=100. These estimates could be lower limits if faster readout patterns are approved in Cycle 2 \citep[see Figure 2 in][]{Batalha2018}. Because of the reset-read pattern, integrations with 3 groups  have an efficiency (duty cycle) of just $(n-1)/(n+1)=50\%$. The most efficient of these patterns has the potential of yielding near $\sim 100\%$ efficiency for bright targets. Additionally, the NIRCam DHS mode is also being considered for Cycle 2+ \citep{Schlawin2017}. If approved, this would enable short wavelength ($1\sim 2$ $\mu$m) $R=300$ spectroscopy on up to $K\sim4.52$ targets. 


\subsection{Additional planets in the \tar\ system?}\label{add_pl}

\par To search for additional potential non-transiting planets in the \tar\ system, we first perform an independent 2pl fit (M3; as listed in Table~\ref{modelcomp}). We place a wide log-uniform prior on the period $P$ between $10^{2}$\ d and $10^{4}$\ d along with a wide uniform prior on the RV semi-amplitude $K$. We find a convergence in the $P-K$ space, implying a possible outer planet companion with period around 3500 d. This periodic signal shows up as a weak peak in the GLS periodogram of the total RV data (See Section \ref{pfs}). This 2pl model has a slight increase $\ln Z$\ ($\Delta \ln Z=\ln Z_{\rm 2pl}-\ln Z_{\rm 1pl}=4.5$). Building on the results from the 2pl model, we perform a further GP+2pl fit (M4) which allows $P$ to vary uniformly between 2000\ d and 4500\ d. We measure $K=2.53\pm0.91$\ m/s for this hypothetical planet. With $P=3279^{+352}_{-307}$\ d, if this signal were indeed caused by a real planet companion, it would correspond to a sub-Saturn planet at $3.9^{+0.3}_{-0.3}$\ AU with $M_{p}\sin i=0.5^{+0.2}_{-0.2}\ M_{\rm Sat}$. However, the increase in the Bayesian evidence from the GP+1pl model, $\ln Z_{\rm GP+2pl}-\ln Z_{\rm GP+1pl}$, is only about 1.0, and thus we conclude that there is insufficient evidence for such a planet in the current RV data.

Such a conclusion is perhaps not surprising, given that the HARPS and PFS data do not have any significant overlap in time over the 20-year baseline that they span, and the proposed planet's period is around 10 years. In addition, it would be extremely challenging to mitigate the stellar activity signals arising from a $\sim 30$~d stellar rotation with an amplitude around $5$~m/s over the course of decades to detect a planet with a considerably smaller RV semi-amplitude ($K\sim 2.5$~m/s). To make things worse, we cannot rule out that the $\sim 3500$ d signal is from the long-term magnetic cycles of the host star, similar to the 11-year solar magnetic cycle. Nonetheless, given the $3\sigma$ constraint on $K$ and $P$ above and the time baseline of our RV data, we have ruled out the existence of an outer gas giant planet with mass down to Saturn mass and period up to roughly 10 years. Future RV follow-up observations with high precision and high cadence could measure the mass of \can\ and might be able to reveal additional planets in the system.\\


\section{Summary and conclusions}\label{conclusions}
In this paper, we characterize the \tar\ planetary system using the publicly available \tess,  HARPS and PFS data in D19 combined with an additional set of more recent PFS RVs. We apply GP models in our transit and RV fits to reduce the effect of stellar rotation and improve the accuracy of the mass determination of \tar b. The sub-Neptune \tar b has a radius of $2.86\pm0.20\ R_{\oplus}$ and an orbital period of $35.6133\pm0.0005$ d. Our GP+Keplerian fit to the RV data reveals that \tar b has a mass of $20.0\pm2.7\ M_{\oplus}$, which agrees with the estimates in D19 within 1$\sigma$. For \can, an Earth-sized planet on a $7.79$ d orbit, we constrain its radius to be $1.13\pm0.10\ R_{\oplus}$ and its mass to have a 3$\sigma$ upper limit at $3.5\ M_{\oplus}$. Compared with D19: (1) we obtain a slightly larger planet radius for both \tar b and \can\ as we apply an updated value of the dilution factor for the \tess\ photometry; and (2) we find a smaller RV semi-amplitude for \tar b after taking the stellar activity into consideration. 

Using Monte Carlo simulations, we verify that our incorporation of a GP model in the RV data analysis does not bias the mass measurement of \tar b. On the other hand, our simulations reveal that a simple Keplerian fit without mitigating stellar activity signals is more likely to result in a biased measurement with an underestimated error bar for \tar b's RV semi-amplitude. Thus we encourage communities to use the updated values in the future as the results in this work are more robust.

The story of \tar b represents a relatively challenging case for RV characterization in several ways. First, it has the inconvenient entanglement of the planetary signal and the stellar activity, which have very similar timescales and exhibit similar RV amplitudes. Additionally, both the stellar rotation and the planet's orbital period are a bit longer than the lunar cycle, which makes it difficult to sample the rotation phase and the orbital phase continuously in complete cycles given that RV spectroscopic observations are normally scheduled during bright time. The fact that we can successfully retrieve the planet's RV semi-amplitude without significant bias illustrates the importance of nightly, high-cadence observations and proper modeling of stellar activity for such systems (similar to \citealt{Morales2016}).

\section*{Acknowledgements}
We thank Nestor Espinoza for his insights and advice on our modeling.
Funding for the \tess\ mission is provided by NASA's Science Mission directorate. 
We acknowledge the use of \tess\ Alert data from pipelines at the \tess\ Science Office and at the \tess\ Science Processing Operations Center. 
Resources supporting this work were provided by the NASA High-End Computing (HEC) Program through the NASA Advanced Supercomputing (NAS) Division at Ames Research Center for the production of the SPOC data products.
This work is partly supported by the National Science Foundation of China (grant No. 11390372 and 11761131004 to S.M. and G.T.J.). 
This research uses data obtained through the Telescope Access Program (TAP), which has been funded by the TAP member institutes.
D. D. acknowledges support from the TESS Guest Investigator Program grant 80NSSC19K1727 and NASA Exoplanet Research Program grant 18-2XRP18\_2-0136.
Support for this work was provided by NASA through Hubble Fellowship grant HSTHF2-51399.001 awarded by the Space Telescope Science Institute, which is operated by the Association of Universities for Research in Astronomy, Inc., for NASA, under contract NAS5-26555. 
Support for this work was also provided by NASA through grant 18-XRP18\_2-0048.
A.D.F. acknowledges the support from the National Science Foundation Graduate Research Fellowship Program under Grant No. (DGE-1746045). Any opinions, findings, and conclusions or recommendations expressed in this material are those of the author(s) and do not necessarily reflect the views of the National Science Foundation.
WH acknowledges funding support of this work through NASA NNH18ZDA001N-XRP grant 80NSSC19K0290. NL, HC, JR, and AG acknowledge funding support by the National Science Foundation CAREER grant 1555175, and the Research Corporation Scialog grants 23782 and 23822. HC is supported by the National Science Foundation Graduate Research Fellowship under Grant No. DGE-1144081. The Evryscope was constructed under National Science Foundation/ATI grant AST-1407589.
Part of this research was carried out at the Jet Propulsion Laboratory, California Institute of Technology, under a contract with the National Aeronautics and Space Administration (80NM0018D0004).
This research has made use of the Exoplanet Follow-up Observation Program website, which is operated by the California Institute of Technology, under contract with the National Aeronautics and Space Administration under the Exoplanet Exploration Program. 
This paper includes data collected by the \tess\ mission, which are publicly available from the Mikulski Archive for Space Telescopes\ (MAST). 
This research made use of observations from Evryscope-South, ESO: 3.6m\ (HARPS) and Magellan: 6.5m\ (PFS).

\section*{Data Availability}
This paper includes photometric data collected by the \tess\ mission, which is publicly available from the Mikulski Archive for Space Telescopes (MAST) at the Space Telescope Science Institure (STScI). The \tess\ light curve used in this work is available in ExoFOP, at \url{https://exofop.ipac.caltech.edu/tess/target.php?id=279741379}. All spectroscopy data underlying this article are listed in the appendix. 

\appendix

\section{GP+Keplerian (GP + 1pl) analysis for PFSS-only data}
\begin{table*}
    \centering
    \caption{Model parameters of \tar b and the best-fit values in the GP+Keplerian (GP + 1pl) analysis for PFSS-only data.}
    \begin{tabular}{lccr}
        \hline\hline
        Parameter       &Best-fit Value       &Prior     &Description\\\hline
        \it{Planetary parameters}\\
        $\rm P_{b}$ (days)   &$35.613$ 
        &Fixed
        &Orbital period of \tar b.\\
        $\rm T_{0,b}$ (BJD)    &$2458350.312$ 
        &Fixed
        &Mid-transit time of \tar b.\\
        $e_{b}\sin \omega_{b}$ &$0.108^{+0.131}_{-0.156}$  &$\mathcal{U}$ ($-1$\ ,\ $1$)  &Parametrisation for $e$ and $\omega$ of \tar b.\\
        $e_{b}\cos \omega_{b}$  &$-0.064^{+0.196}_{-0.187}$ &$\mathcal{U}$ ($-1$\ ,\ $1$)  &Parametrisation for $e$ and $\omega$ of \tar b.\\
        \\
        \it{RV parameters}\\
        $\rm \mu_{PFSS}$ ($\rm m\ s^{-1}$)    &$0.66^{+1.51}_{-1.53}$ 
        &$\mathcal{U}$ ($-10$\ ,\ $10$)
        &RV offset for post-upgrade PFS (PFSS).\\
        $\rm \sigma_{PFSS}$ ($\rm m\ s^{-1}$)    &$0.28^{+0.17}_{-0.16}$ 
        &$\mathcal{U}$ ($0$\ ,\ $10$)
        &Jitter term for post-upgrade PFS (PFSS).\\
        $K_{b}$ ($\rm m\ s^{-1}$)       &$4.94^{+1.18}_{-1.17}$
        &$\mathcal{N}$ ($5.5$\ ,\ $2^{2}$)
        &RV semi-amplitude of \tar\ b.\\
        
        \\
        \it{GP hyperparameters}\\
        $B$ ($\rm (m\ s^{-1})^{2}$)  &$19.442^{+7.627}_{-4.964}$
        &$\mathcal{J}$ ($10^{-3}$\ ,\ $10^{3}$)      &Covariance amplitude of the global GP component.\\
        $C$  &$0.037^{+0.478}_{-0.033}$
        &$\mathcal{J}$ ($10^{-3}$\ ,\ $10^{3}$)      &Factor of the global GP component.\\
        $L$ (days)  &$27.693^{+10.563}_{-5.271}$
        &$\mathcal{N}$ ($57.8$\ ,\ $50^{2}$)      &Exponential
        evolutionary timescale of the global GP component.\\
        $P_{\rm rot}$ (days)  &$32.4^{+4.1}_{-3.6}$
        &$\mathcal{N}$ ($31$\ ,\ $4^{2}$)      &Periodic
        timescale of the global GP component.\\
        
         \hline\hline 
    \end{tabular}
    \label{rvgppriors_PFSS}
\end{table*}

\section{All RVs and stellar activity indicators of HARPS and PFS}
\begin{table*}
    \centering
    \caption{HARPS RVs and stellar activity indicators}
    \begin{tabular}{ccccccc}
        \hline\hline
        Time(BJD)   &RV(m/s)  &RVerr(m/s)     &CRX &CRXerr &DLW  &DLWerr   \\\hline
        2452944.721	&-1.34 	&0.31 	&-8.81 	&4.79 	&-3.68 	&1.31\\ 
        2452999.632	&-1.16 	&0.43 	&-7.72 	&5.31 	&-23.34 &1.44\\ 
        2453229.926	&-7.89 	&0.32 	&-1.85 	&3.03 	&-14.83 &1.11\\
        2453269.827	&-6.83 	&0.56 	&8.24 	&4.45 	&-2.86 	&1.59\\ 
        2453272.829	&-7.28 	&0.33 	&7.95 	&3.16 	&-17.44 &1.22\\ 
        2453668.802	&-0.01 	&0.39 	&3.61 	&3.72 	&-22.05 &1.10\\ 
        2453765.524	&-11.93 &0.23 	&10.00 	&3.03 	&-30.58 &1.43\\
        2453786.575	&5.13 	&0.32 	&3.72 	&3.59 	&-22.65 &1.23\\ 
        2453787.561	&6.61 	&0.31 	&4.64 	&3.09 	&-21.81 &1.20\\ 
        2453974.883	&-5.89 	&0.39 	&-3.97 	&3.62 	&-10.69 &0.91\\ 
        2453980.853	&-4.98 	&0.33 	&1.07 	&3.70 	&-10.63 &1.08\\ 
        2453983.878	&-7.53 	&0.26 	&1.32 	&2.86 	&-18.41 &1.00\\ 
        2454047.702	&-10.84 &0.31 	&11.24 	&3.36 	&-29.74 &1.06\\ 
        2454049.705	&-6.76 	&0.31 	&2.57 	&4.56 	&-23.58 &0.94\\ 
        2454051.726	&-2.64 	&0.34 	&5.83 	&2.91 	&-17.83 &0.86\\ 
        2454083.692	&-6.69 	&0.32 	&9.08 	&3.23 	&-24.62 &0.90\\
        2454120.634	&-13.24 &0.36 	&5.01 	&3.21 	&-2.39 	&2.28\\ 
        2454121.598	&-11.41 &0.38 	&2.45 	&4.54 	&-26.87 &1.18\\
        2454314.887	&2.58 	&0.43 	&7.45 	&4.08 	&5.91 	&0.85\\ 
        2454316.899	&0.01 	&0.30 	&-4.08 	&3.95 	&2.52 	&0.70\\ 
        2454319.876	&-2.22 	&0.38 	&-5.89 	&4.15 	&-5.78 	&1.00\\ 
        2454528.522	&8.31 	&0.30 	&2.81 	&4.19 	&-4.05 	&0.59\\ 
        2454705.921	&2.68 	&0.75 	&1.06 	&7.04 	&12.38 	&1.52\\ 
        2454706.899	&5.52 	&1.15 	&0.97 	&10.95 	&20.36 	&1.86\\ 
        2454707.903	&5.78 	&0.36 	&-5.09 	&4.06 	&5.36 	&0.88\\ 
        2454710.887	&8.03 	&0.37 	&-1.72 	&4.05 	&6.18 	&0.92\\ 
        2454732.763	&-1.56 	&0.38 	&11.30 	&4.26 	&-11.82	&0.85\\ 
        2454736.806	&4.84 	&0.40 	&4.17 	&4.06 	&10.37 	&0.81\\ 
        2454738.849	&4.32 	&0.62 	&9.14 	&5.51 	&16.83 	&1.49\\ 
        2455025.910	&4.28 	&0.39 	&4.66 	&4.44 	&30.61 	&1.31\\ 
        2455040.898	&3.24 	&0.72 	&-17.95 &7.64 	&45.73 	&1.71\\ 
        2455041.897	&-2.19 	&0.37 	&-17.80 &3.67 	&36.22 	&1.15\\ 
        2455043.900	&-4.94 	&0.51 	&-15.84 &4.77 	&26.34 	&0.97\\ 
        2455045.888	&-10.79 &0.40 	&2.92 	&4.13 	&11.03 	&0.86\\ 
        2455046.847	&-10.30 &0.41 	&-4.47 	&4.96 	&5.62 	&0.86\\ 
        2455067.891	&7.31 	&0.26 	&3.39 	&3.60 	&15.37 	&0.97\\ 
        2455068.893	&9.87 	&0.38 	&-2.65 	&4.35 	&22.54 	&1.13\\ 
        2455070.906	&9.67 	&0.30 	&4.14 	&3.51 	&32.70 	&1.39\\
        2455073.884	&11.14 	&0.30 	&-1.26 	&3.84 	&44.13 	&1.39\\ 
        2455075.872	&6.72 	&0.37 	&-16.64 &3.87 	&50.61 	&1.66\\ 
        2455103.812	&1.93 	&0.51 	&-1.68 	&4.78 	&19.72 	&1.20\\ 
        2455108.791	&4.44 	&0.42 	&3.61 	&3.63 	&21.62 	&1.15\\ 
        2455110.770	&4.17 	&0.89 	&8.21 	&8.08 	&39.61 	&1.85\\ 
        2455112.790	&2.58 	&0.31 	&-6.26 	&3.81 	&26.84 	&1.00\\ 
        2455115.784	&-3.03 	&0.39 	&-11.45 &4.39 	&27.77 	&1.01\\ 
        2455122.775	&-5.81 	&0.36 	&-8.17 	&3.98 	&2.36 	&0.75\\ 
        2455124.781	&-4.63 	&0.24 	&1.80 	&3.17 	&-2.51 	&0.56\\ 
        2455128.793	&-1.94 	&0.32 	&0.08 	&3.45 	&-9.62 	&0.71\\ 
        2455133.765	&5.24 	&0.38 	&-0.69 	&4.31 	&4.99 	&0.81\\ 
        2455137.765	&1.77 	&0.27 	&-6.96 	&3.22 	&2.45 	&0.63\\ 
        2455141.675	&2.71 	&0.30 	&0.54 	&3.00 	&1.56 	&0.66\\ 
        2455164.591	&-0.01 	&0.26 	&-2.67 	&2.87 	&6.32 	&0.54\\ 
        2455167.579	&3.62 	&0.24 	&-4.53 	&2.96 	&0.17 	&0.61\\ 
        2455169.584	&3.56 	&0.27 	&-4.67 	&2.72 	&3.03 	&0.61\\ 
        2457741.706	&0.39 	&0.85 	&11.10 	&5.43 	&-6.29 	&0.43\\ 
        2457743.716	&-1.79 	&0.82 	&-4.01 	&3.86 	&0.19 	&0.37\\ 
        2457744.741	&-2.08 	&0.85 	&-5.14 	&3.75 	&3.68 	&0.34\\ 
        2457747.711	&-7.32 	&0.85 	&-6.96 	&7.55 	&7.98 	&0.70\\ 
        \hline\hline 
    \end{tabular}
    \label{harpsai}
\end{table*}

\begin{table}
    \centering
    \caption{PFS RVs and stellar activity indicators}
    \begin{tabular}{ccccc}
        \hline\hline
        Time (BJD)   &RV (m/s)  &RVerr (m/s)     &$S_{\rm HK}$ &$S_{\rm H\alpha}$   \\\hline
2455198.63 &2.57 &1.01 &0.3248 &0.0448 \\
2455198.632 &1.12 &1.06 &0.3323 &0.0446 \\
2455430.896 &5.98 &0.98 &0.3078 &0.0436 \\
2455586.569 &1.78 &0.88 &0.3021 &0.043 \\
2455785.851 &-5.61 &0.94 &0.2822 &0.0432 \\
2455790.843 &-5.72 &1.13 &0.3158 &0.0434 \\
2455793.883 &-4.14 &1.2 &0.3166 &0.0439 \\
2455795.922 &-4.12 &1.21 &0.3041 &0.0446 \\
2456141.926 &5.18 &1.14 &0.3 &0.0436 \\
2456284.62 &3.3 &0.79 &0.4561 &0.0431 \\
2456293.63 &-4.64 &0.78 &0.2864 &0.0433 \\
2456345.527 &4.01 &1.28 &0.379 &0.0431 \\
2456501.925 &-1.79 &1.09 &0.31 &0.0431 \\
2456548.858 &0.56 &1.07 &0.353 &0.043 \\
2456551.837 &-0.14 &1.67 &0.3521 &-1.0 \\
2456556.829 &5.31 &1.05 &0.2892 &0.0444 \\
2456604.679 &-2.71 &0.81 &0.3013 &0.0433 \\
2456607.721 &-10.12 &0.92 &0.2989 &0.0433 \\
2456611.744 &-4.62 &0.89 &0.3015 &0.044 \\
2456612.677 &-3.89 &0.89 &0.2927 &0.0434 \\
2456692.543 &1.5 &0.95 &0.3021 &0.0431 \\
2456694.559 &0.22 &0.94 &0.2835 &0.044 \\
2456698.547 &-0.43 &1.06 &0.3025 &0.043 \\
2456866.897 &-2.74 &1.15 &0.2999 &0.043 \\
2456871.905 &-1.55 &1.03 &0.2838 &0.0432 \\
2456877.917 &2.74 &1.15 &0.3891 &0.0432 \\
2457022.633 &-3.35 &0.77 &0.2926 &0.0432 \\
2457027.631 &-0.38 &0.83 &0.2914 &0.0441 \\
2457052.553 &-2.92 &0.88 &0.294 &0.0432 \\
2457064.54 &-3.2 &1.0 &0.3909 &0.0434 \\
2457260.897 &-0.05 &1.07 &0.3124 &0.0426 \\
2457267.853 &-0.25 &1.04 &0.302 &0.0434 \\
2457319.787 &-5.37 &0.93 &0.2985 &0.0433 \\
2457323.797 &-7.37 &0.93 &0.3067 &0.0435 \\
2457389.609 &-7.63 &0.88 &0.2894 &0.0432 \\
2457616.915 &6.98 &1.09 &0.2939 &0.0432 \\
2457626.922 &10.73 &1.08 &0.2881 &0.0439 \\
2457737.648 &4.36 &0.84 &0.3004 &0.0428 \\
2457740.613 &9.51 &0.82 &0.2833 &0.0426 \\
2457759.556 &-0.92 &0.8 &0.2854 &0.0434 \\
2457764.59 &11.8 &0.82 &0.3011 &0.0446 \\
2457768.579 &8.34 &0.78 &0.4384 &0.043 \\
2458356.871 &-5.48 &0.92 &0.4683 &0.044 \\
2458356.875 &-3.03 &0.88 &0.4652 &0.0436 \\
2458407.77 &3.76 &0.8 &0.4698 &0.0438 \\
2458411.721 &5.37 &0.85 &0.5125 &0.0442 \\
2458411.725 &5.77 &0.86 &0.5218 &0.0442 \\
2458417.721 &7.72 &0.85 &0.4847 &0.0435 \\
2458467.651 &-5.51 &0.76 &0.4517 &0.0437 \\
2458467.655 &-5.26 &0.71 &0.4517 &0.0438 \\
2458467.783 &-5.5 &0.78 &0.4466 &0.0432 \\
2458467.787 &-3.34 &0.75 &0.4498 &0.0432 \\
2458468.626 &-2.48 &0.8 &0.4711 &0.0439 \\
2458468.63 &-1.63 &0.8 &0.4622 &0.0433 \\
2458469.625 &-0.35 &0.77 &0.4671 &0.044 \\
2458469.629 &0.87 &0.72 &0.4664 &0.0435 \\
2458469.751 &1.03 &0.74 &0.4733 &0.0434 \\
2458469.755 &-0.03 &0.74 &0.47 &0.0434 \\
2458471.627 &1.91 &0.7 &0.4681 &0.0436 \\
2458471.632 &1.1 &0.73 &0.4598 &0.0435 \\
2458471.676 &1.16 &0.75 &0.471 &0.0439 \\
2458471.68 &2.36 &0.69 &0.4692 &0.0439 \\
2458472.606 &1.16 &0.71 &0.4577 &0.0439 \\
2458472.61 &1.78 &0.71 &0.4528 &0.044 \\

        \hline\hline 
    \end{tabular}
    \label{pfsai}
\end{table}

\begin{table}
 \contcaption{PFS RVs and stellar activity indicators}
 \label{tab:continued}
 \begin{tabular}{ccccc}
  \hline\hline
  Time (BJD)   &RV (m/s)  &RVerr (m/s)     &$S_{\rm HK}$ &$S_{\rm H\alpha}$   \\\hline
  2458472.744 &-0.67 &0.73 &0.4651 &0.0433 \\
2458472.748 &2.74 &0.69 &0.4856 &0.0432 \\
2458473.573 &2.41 &0.73 &0.4722 &0.0436 \\
2458473.577 &2.59 &0.68 &0.4666 &0.0435 \\
2458473.644 &2.1 &0.67 &0.4589 &0.0435 \\
2458473.648 &2.57 &0.71 &0.4567 &0.0435 \\
2458474.655 &2.24 &0.75 &0.4628 &0.0439 \\
2458474.659 &2.26 &0.65 &0.4545 &0.0441 \\
2458474.73 &1.37 &0.79 &0.4861 &0.0438 \\
2458474.734 &2.25 &0.83 &0.4906 &0.0439 \\
2458475.587 &2.31 &0.68 &0.4641 &0.0435 \\
2458475.591 &3.17 &0.69 &0.4668 &0.0436 \\
2458475.715 &2.43 &0.7 &0.4658 &0.0435 \\
2458475.72 &2.28 &0.74 &0.4712 &0.0434 \\
2458476.582 &2.98 &0.73 &0.4705 &0.0438 \\
2458476.586 &3.98 &0.68 &0.4685 &0.0437 \\
2458476.674 &3.56 &0.65 &0.4646 &0.0441 \\
2458476.678 &3.88 &0.67 &0.463 &0.0441 \\
2458479.607 &9.91 &0.66 &0.4783 &0.0436 \\
2458479.611 &9.78 &0.66 &0.4784 &0.0437 \\
2458479.747 &10.16 &0.78 &0.5004 &0.0436 \\
2458479.751 &9.1 &0.76 &0.513 &0.0437 \\
2458480.637 &9.07 &0.78 &0.4949 &0.0437 \\
2458480.649 &9.49 &0.73 &0.4956 &0.0436 \\
2458480.735 &11.96 &0.75 &0.4928 &0.0435 \\
2458480.739 &10.72 &0.77 &0.491 &0.0435 \\
2458501.532 &-8.46 &0.7 &0.4341 &0.0428 \\
2458501.536 &-7.72 &0.81 &0.4452 &0.0428 \\
2458501.61 &-9.73 &0.74 &0.4418 &0.0428 \\
2458501.614 &-7.92 &0.73 &0.4412 &0.0428 \\
2458502.522 &-9.63 &0.75 &0.431 &0.0435 \\
2458502.526 &-8.32 &0.73 &0.4293 &0.0434 \\
2458502.643 &-8.86 &0.75 &0.4298 &0.0427 \\
2458502.648 &-9.64 &0.74 &0.4395 &0.0428 \\
2458503.538 &-7.62 &0.77 &0.4269 &0.0434 \\
2458503.629 &-7.28 &0.71 &0.4408 &0.043 \\
2458503.633 &-7.18 &0.74 &0.4278 &0.0429 \\
2458504.623 &-7.13 &0.74 &0.4335 &0.0428 \\
2458504.627 &-6.97 &0.72 &0.4272 &0.0429 \\
2458504.679 &-7.13 &0.74 &0.4373 &0.0429 \\
2458504.683 &-5.61 &0.7 &0.4467 &0.043 \\
2458505.525 &-6.11 &0.72 &0.4303 &0.043 \\
2458505.529 &-7.0 &0.73 &0.4365 &0.043 \\
2458505.613 &-6.59 &0.76 &0.4448 &0.043 \\
2458505.617 &-6.66 &0.83 &0.4828 &0.043 \\
2458506.523 &-5.71 &0.76 &0.4496 &0.0435 \\
2458506.527 &-5.03 &0.88 &0.4612 &0.0433 \\
2458506.595 &-5.46 &0.82 &0.4818 &0.0433 \\
2458506.6 &-2.99 &0.92 &0.4843 &0.0432 \\
2458507.527 &-2.57 &0.86 &0.4613 &0.0434 \\
2458507.531 &-2.37 &0.87 &0.4851 &0.0434 \\
2458507.601 &-4.29 &0.67 &0.4231 &0.043 \\
2458508.523 &-1.87 &0.74 &0.4489 &0.0428 \\
2458508.528 &-3.87 &0.84 &0.4415 &0.0428 \\
2458508.652 &-2.17 &0.93 &0.4681 &0.0432 \\
2458508.657 &-2.91 &0.83 &0.4851 &0.0432 \\
2458509.523 &-1.25 &0.89 &0.4762 &0.0432 \\
2458509.527 &-1.49 &0.87 &0.4796 &0.0431 \\
2458510.524 &1.34 &1.02 &0.4926 &0.0433 \\
2458510.528 &0.62 &1.02 &0.5029 &0.0432 \\
2458511.521 &0.06 &0.79 &0.4574 &0.0436 \\
2458511.525 &-0.66 &0.77 &0.4573 &0.0435 \\
2458526.526 &-5.53 &0.96 &0.5108 &0.0431 \\
2458526.53 &-2.79 &0.86 &0.4894 &0.0429 \\

  \hline\hline
 \end{tabular}
\end{table}

\begin{table}
 \contcaption{PFS RVs and stellar activity indicators}
 \label{tab:continued}
 \begin{tabular}{ccccc}
  \hline\hline
    Time (BJD)   &RV (m/s)  &RVerr (m/s)     &$S_{\rm HK}$ &$S_{\rm H\alpha}$   \\\hline
  2458527.514 &-3.71 &1.04 &0.4926 &0.0435 \\
  2458527.519 &-5.36 &0.98 &0.5053 &0.0434 \\
2458527.647 &-0.58 &1.19 &0.5673 &0.0428 \\
2458527.652 &-2.91 &1.15 &0.5594 &0.0424 \\
2458528.516 &-4.36 &0.9 &0.4705 &0.0434 \\
2458528.52 &-1.56 &0.91 &0.4669 &0.0434 \\
2458528.583 &-3.25 &0.94 &0.4535 &0.0435 \\
2458528.587 &-4.76 &1.02 &0.489 &0.0434 \\
2458529.521 &-3.44 &0.92 &0.4432 &0.0429 \\
2458529.525 &-4.82 &0.95 &0.4466 &0.043 \\
2458529.599 &-4.89 &0.9 &0.4597 &0.0433 \\
2458529.603 &-3.59 &0.93 &0.4628 &0.0433 \\
2458530.511 &-2.51 &0.98 &0.4602 &0.0428 \\
2458530.515 &-5.21 &0.92 &0.4546 &0.0429 \\
2458530.585 &-7.03 &0.92 &0.4452 &0.0428 \\
2458530.59 &-5.35 &0.88 &0.4662 &0.0427 \\
2458531.604 &-5.22 &0.87 &0.4642 &0.0424 \\
2458531.609 &-5.12 &0.88 &0.4789 &0.0424 \\
2458532.513 &-5.5 &0.89 &0.4558 &0.0431 \\
2458532.517 &-3.41 &0.89 &0.4712 &0.0429 \\
2458532.604 &-5.96 &0.96 &0.4772 &0.0429 \\
2458532.608 &-4.94 &0.9 &0.4657 &0.0429 \\
2458674.921 &-5.03 &0.89 &0.4512 &0.044 \\
2458676.921 &-5.15 &1.16 &0.4791 &0.0435 \\
2458676.928 &-5.0 &0.93 &0.4381 &0.0437 \\
2458677.893 &-4.72 &0.99 &0.4588 &0.0441 \\
2458677.897 &-2.52 &1.0 &0.4617 &0.0441 \\
2458678.892 &-4.82 &0.92 &0.4558 &0.044 \\
2458678.896 &-2.29 &0.99 &0.4536 &0.0441 \\
2458681.894 &2.27 &1.18 &0.5196 &0.0442 \\
2458681.898 &1.65 &0.99 &0.4938 &0.0445 \\
2458683.905 &1.99 &0.95 &0.46 &0.0433 \\
2458683.914 &-0.32 &0.92 &0.4505 &0.0439 \\
2458684.869 &-2.07 &0.85 &0.4488 &0.0439 \\
2458684.873 &-1.68 &0.88 &0.4473 &0.0439 \\
2458709.922 &-0.56 &0.9 &0.45 &0.0436 \\
2458709.927 &-1.05 &0.91 &0.4488 &0.0437 \\
2458737.78 &7.85 &1.14 &0.4805 &0.0437 \\
2458737.873 &9.19 &1.1 &0.4961 &0.0436 \\
2458737.877 &7.22 &1.18 &0.5039 &0.0438 \\
2458738.75 &2.55 &0.97 &0.4954 &0.0438 \\
2458738.754 &3.93 &1.13 &0.4903 &0.0438 \\
2458738.88 &3.64 &1.09 &0.4989 &0.0441 \\
2458738.883 &2.18 &1.12 &0.5169 &0.0442 \\
2458738.888 &3.6 &1.1 &0.517 &0.0441 \\
2458739.776 &2.97 &0.97 &0.482 &0.0441 \\
2458739.869 &1.24 &0.98 &0.4715 &0.0436 \\
2458740.776 &1.91 &0.95 &0.4771 &0.0437 \\
2458740.862 &0.29 &0.94 &0.4793 &0.0437 \\
2458741.782 &-0.62 &0.99 &0.4868 &0.0435 \\
2458741.785 &1.9 &1.04 &0.4875 &0.0436 \\
2458744.796 &0.54 &1.06 &0.4922 &0.0435 \\
2458744.799 &-1.16 &1.1 &0.4903 &0.0435 \\
2458744.874 &-0.94 &1.02 &0.4625 &0.0435 \\
2458760.775 &0.3 &0.87 &0.4255 &0.0441 \\
2458760.846 &1.18 &0.91 &0.4308 &0.044 \\
2458761.847 &-0.92 &0.87 &0.4285 &0.0435 \\
2458762.762 &-0.67 &0.94 &0.4483 &0.0433 \\
2458762.767 &0.94 &0.97 &0.4491 &0.0435 \\
2458762.838 &0.0 &0.96 &0.4431 &0.0434 \\
2458762.842 &-1.69 &0.92 &0.4394 &0.0435 \\
2458763.804 &1.27 &0.8 &0.406 &0.044 \\
2458764.766 &2.41 &0.92 &0.4494 &0.0438 \\
2458764.77 &3.0 &0.89 &0.4498 &0.0439 \\
  \hline\hline
 \end{tabular}
\end{table}

\begin{table}
 \contcaption{PFS RVs and stellar activity indicators}
 \label{tab:continued}
 \begin{tabular}{ccccc}
  \hline\hline
  Time (BJD)   &RV (m/s)  &RVerr (m/s)     &$S_{\rm HK}$ &$S_{\rm H\alpha}$   \\\hline
2458764.837 &2.53 &0.91 &0.4642 &0.0433 \\
2458764.842 &2.9 &0.93 &0.4482 &0.0434 \\
2458766.757 &7.0 &0.84 &0.4269 &0.0436 \\
2458766.839 &5.87 &0.78 &0.4332 &0.0433 \\
2458767.752 &8.35 &0.86 &0.4398 &0.0441 \\
2458768.764 &8.85 &0.89 &0.4857 &0.0442 \\
2458768.839 &10.45 &0.87 &0.4452 &0.0437 \\
2458828.579 &5.06 &0.71 &0.4599 &0.0433 \\
2458828.586 &6.45 &0.74 &0.4904 &0.0433 \\
2458828.657 &5.66 &0.72 &0.5039 &0.0434 \\
2458828.661 &5.7 &0.78 &0.4898 &0.0434 \\
2458829.726 &7.66 &0.9 &0.4901 &0.0433 \\
2458829.73 &8.43 &0.9 &0.5023 &0.0434 \\
2458831.616 &7.26 &0.83 &0.4886 &0.0433 \\
2458831.621 &7.08 &0.75 &0.4865 &0.0434 \\
2458831.702 &6.87 &0.81 &0.4915 &0.0433 \\
2458831.705 &6.82 &0.78 &0.5013 &0.0434 \\
2458832.584 &5.69 &0.89 &0.5113 &0.0428 \\
2458832.589 &7.38 &0.82 &0.4735 &0.043 \\
2458832.625 &6.62 &0.98 &0.4884 &0.0432 \\
2458832.632 &5.63 &0.91 &0.4933 &0.0431 \\
2458833.621 &5.57 &0.84 &0.4904 &0.0433 \\
2458833.625 &3.27 &0.87 &0.4919 &0.0433 \\
2458833.689 &6.29 &0.91 &0.4978 &0.0431 \\
2458833.693 &6.03 &0.79 &0.4841 &0.0432 \\
2458834.603 &2.7 &0.83 &0.4851 &0.0434 \\
2458834.606 &3.74 &0.83 &0.4849 &0.0435 \\
2458834.677 &3.58 &0.83 &0.4732 &0.0435 \\
2458834.681 &2.03 &0.84 &0.4549 &0.0433 \\
2458837.651 &0.38 &0.85 &0.4765 &0.0434 \\
2458837.655 &0.09 &0.85 &0.4666 &0.0436 \\
2458887.535 &-6.62 &1.06 &0.4883 &0.0443 \\
2458887.539 &-5.47 &0.92 &0.5091 &0.0442 \\
  \hline\hline
 \end{tabular}
\end{table}



\bibliographystyle{mnras}
\bibliography{planet} 

\bsp	
\label{lastpage}
\end{document}